\documentclass[twocolumn,showpacs,preprintnumbers,amsmath,amssymb]{revtex4}
\newcommand{\LL}{\mathcal{L}}
\newcommand{\HH}{\mathcal{H}}
\newcommand*{\simu}[1]{\widetilde{#1}}

\begin{document}

\title{Superselection rules and quantum protocols}

\author{Alexei Kitaev,$^1$ Dominic Mayers,$^{1,2}$ and John Preskill$^1$}
\affiliation{$^1$ Institute for Quantum Information, California Institute of Technology, 
Pasadena, CA 91125, USA\\
$^2$ D\'epartement de Math\'ematiques et d'Informatique, Universit\'e de Sherbrooke, Qu\'ebec, Canada}

\begin{abstract}
We show that superselection rules do not enhance the information-theoretic security of quantum cryptographic protocols. Our analysis employs two quite different methods. The first method uses the concept of a {\em reference system} --- in a world subject to a superselection rule, unrestricted operations can be simulated by parties who share access to a reference system with suitable properties. By this method, we prove that if an $n$-party protocol is secure in a world subject to a superselection rule, then the security is maintained even if the superselection rule is relaxed. However, the proof applies only to a limited class of superselection rules, those in which the superselection sectors are labeled by unitary irreducible representations of a compact symmetry group. The second  method uses the concept of the {\em format} of a message sent between parties --- by verifying the format, the recipient of a message can check whether the message could have been sent by a party who performed charge-conserving operations. By this method, we prove that protocols subject to general superselection rules (including those pertaining to nonabelian anyons in two dimensions) are no more secure than protocols in the unrestricted world. However, the proof applies only to two-party protocols. Our results show in particular that, if no assumptions are made about the computational power of the cheater, then secure quantum bit commitment and strong quantum coin flipping with arbitrarily small bias are impossible in a world subject to superselection rules. 

\end{abstract}

\maketitle

\section{Introduction}
\label{sec:intro}
The central aim of  modern cryptography is to formulate protocols that achieve cryptographic tasks with {\em computational security}, meaning that a dishonest party would need to perform a prohibitively difficult computation to break the protocol. A major goal of quantum cryptography is to formulate protocols, involving the exchange of quantum states, that achieve {\em information-theoretic security}, meaning that even an adversary with unlimited computational power would be unable to defeat the protocol \cite{BennettBrassard84}. Information-theoretic security (sometimes called ``unconditional security'') has been established for quantum key distribution protocols \cite{Mayers96,LoChau99,BBBMR00,ShorPreskill00,KoashiPreskill02,Koashi03} but it has also been shown that, even in the quantum world, information-theoretic security is not attainable for certain tasks. For example, unconditionally secure quantum bit commitment is impossible \cite{mayers_qbc,lo_chau}, as is (strong) quantum coin flipping with arbitrarily small bias \cite{kitaev_coin,ambainis_kitaev}.

Superselection rules are limitations on the physically realizable quantum operations that can be carried out by a local agent. For example, it is impossible to create or destroy an isolated particle that carries locally conserved charges, such as an electrically charged particle, a fermion, or (in a two-dimensional medium) an anyon. Recently, Popescu \cite{popescu} has suggested that superselection rules might have interesting implications for the security of quantum cryptographic protocols. The intuitive idea behind this suggestion is that superselection rules could place inviolable limits on the cheating strategies available to the dishonest parties, thus enhancing security. Might, say, unconditionally secure bit commitment be possible in worlds (perhaps including the physical world that we inhabit) governed by suitable superselection rules? An affirmative answer could shake the foundations of cryptography.

The purpose of this paper is to answer Popescu's intriguing question. Sadly, our conclusion is that superselection rules can never foil a cheater who has unlimited quantum-computational power. 

In the case of quantum bit commitment, and other two-party protocols, our argument hinges on a quite simple observation.  In a two-party protocol, one participant (Alice) has control of a local system $A$, and the other participant (Bob) has control of another local system $B$. In addition, there is a message system $M$ that they pass back and forth. In each step of the protocol, one party performs a joint quantum operation on her/his local system and the message system, and then sends the message system to the other party. Suppose that in each step, any part of the full system $ABM$ that is beyond Alice's control is under Bob's control and vice-versa --- no part of the full system is inaccessible or in the possession of a third party.  Suppose further that the full system $ABM$ has trivial total charge (belongs to the trivial superselection sector). Then at any stage of the protocol, the algebra of operations that Alice can perform is the {\em commutant} of the algebra of operations that Bob can perform; that is, Alice's algebra contains {\em all} operations that commute with Bob's algebra. Likewise, Bob's algebra is the commutant of Alice's. By a minor extension of the standard argument, it then follows that unconditionally secure quantum bit commitment is impossible {\em if} the total charge shared by the parties is trivial.

Now, if the total charge in {\em nontrivial}, then Alice's algebra is surely a subalgebra of the commutant of Bob's, but it may be a {\em proper} subalgebra; similarly, Bob's algebra may be a proper subalgebra of Alice's. This unusual property of the local operations seems to open new possibilities for the design of quantum protocols. Regrettably, though, there is no way for an honest party to ensure that the total charge is really nontrivial, when the other party is dishonest. Though the honest protocol may call for the parties to start out with nontrivial charges, we may always imagine that there are actually compensating charges beyond the grasp of Alice and Bob, so that the total charge of the world is really trivial. Furthermore, a  cheater might seize control of the compensating charge, while for an honest party it makes no difference whether the compensating charge is present or not. It follows that a protocol that calls for the total charge to be nontrivial can be no more secure than one in which the total charge is actually trivial; we conclude again that unconditionally secure quantum bit commitment is impossible, irrespective of the value of the total charge shared by the parties in the honest protocol.

Aside from quantum bit commitment, we will also study the impact of superselection rules on the information-theoretic security of a broad class of other quantum protocols, using two different methods. We analyze in detail the important special case where the superselection sectors can be identified with the unitary irreducible representations of a compact symmetry group. In that case, we argue that it is possible in principle to prepare a {\em reference state} that establishes a preferred orientation in the symmetry group. A party with access to the reference state can use it to perform operations that are ostensibly forbidden by the superselection rule. In particular, consider an $n$-party quantum protocol where up to $k<n$ of the parties are dishonest, and suppose that in a world with no superselection rules the dishonest parties have a cheating strategy that breaks the protocol. Then, even in a world with superselection rules, the dishonest parties, by sharing a suitable reference state, can simulate this cheating strategy faithfully. We conclude that if a quantum protocol is information-theoretically secure in a world with a superselection rule, the security will be maintained even if the superselection rule is relaxed, at least in the case where the superselection rule arises from a compact symmetry group.

Superselection rules arising from compact symmetry groups are not the most general possible ones. In particular, an especially rich variety of superselection rules are potentially realizable in two-dimensional systems such as those that admit nonabelian anyons. However even superselection rules of this more general kind cannot foil a cheater. We find that for any two-party protocol that is secure in a world subject to a superselection rule, the security is maintained when the superselection rule is relaxed. 

Our analysis of these more general superselection rules does not rely on the concept of a reference system; rather it is founded on a completely different idea, the concept of the {\em format} of a message.  A superselection rule can always be characterized by saying that there are charges that must be conserved by all local operations, and when we relax the superselection rule, in effect we are permitting a cheater to violate these conservation laws. For the purpose of assessing the security of a two-party protocol, we are interested in how the actions of the cheating party (Alice) affect the outcomes of measurements performed by the honest party (Bob). Potentially, if Alice is granted the power to violate conservation of ``charge,'' her ability to influence Bob's measurements will be strengthened. 

However, if the total charge shared by Alice and Bob is trivial (as we are entitled to assume in an analysis of security), then if charge is conserved, Alice and Bob hold conjugate charges at each stage of the protocol. Therefore, Bob always knows what charge Alice is supposed to have, which constrains the type of message that Alice can send to Bob if she is honest. When Bob receives a message he can verify its format, checking whether the message could have been sent by a party who performed a charge-conserving operation, and he can abort the protocol if the verification fails. Therefore, if the protocol ends normally, Alice has been forced to respect charge conservation --- her power to flout the superselection rule does not enhance her ability to fool Bob. This reasoning shows that superselection rules cannot thwart cheating, but because the argument relies on the property that Alice and Bob hold perfectly correlated charges, it works only for two-party protocols.

For cryptographic protocols with more than two parties, and for general superselection rules, new subtleties arise. In two spatial dimensions, general charges are not merely locally conserved, they may also have nontrivial {\em braiding} properties --- the exchange of two charges may induce a nontrivial transformation on their joint Hilbert space. This means that the effect of sending a message from one party to another can depend on the path along which the message travels. It is an interesting problem to specify appropriate definitions of security for protocols in this setting, but we will not attempt to address this issue here. For the special case of charges labeled by unitary representations of compact groups, the braiding properties are trivial; therefore in that case we can analyze multiparty protocols without confronting such questions.

Verstraete and Cirac \cite{verstraete} recently discussed a data-hiding protocol whose security is premised on a superselection rule. However, as the authors recognized, the protocol is not unconditionally secure; it can be broken if the parties establish a suitable shared reference state via quantum communication. The notion that the naive implications of a superselection rule can be evaded through the use of a suitable reference system was emphasized long ago by Aharonov and Susskind \cite{aharonov}; see \cite{bartlett_ent} for a recent discussion. A special case of our main result was reported earlier in \cite{mayers}. 

The rest of this paper is organized as follows: We develop the concept of a reference system in Sec.~\ref{sec:reference}, first for abelian, then for nonabelian symmetries, and we explain how a reference system can be used to simulate unrestricted operations in a world subject to superselection rules arising from a symmetry group; this observation is applied in Sec.~\ref{sec:simulation} to the analysis of the security of quantum protocols. In Sec.~\ref{sec:distributed} we explore the distinction between an {\em itinerant} reference system that is passed from party to party as needed during a protocol, and a {\em distributed} reference system that can be prepared and passed out to the parties before the protocol begins. Superselection rules arising from nonabelian symmetries are further characterized in Sec.~\ref{sec:commutants}, and we comment in Sec.~\ref{sec:data} on the data-hiding protocol of Verstraete and Cirac. Our analysis of the impact of superselection rules on the security of quantum bit commitment is in Sec.~\ref{sec:qbc}; we also show there that for the analysis of security of an $n$-party protocol, it suffices to consider the case in which the total charge held by the parties is trivial. Two-party protocols subject to general superselection rules are investigated in Sec.~\ref{sec:two-party}, and Sec.~\ref{sec:conclusions} contains some concluding comments.

\section{Superselection rules and reference systems}
\label{sec:reference}
A superselection rule is a decomposition of Hilbert space into sectors that are preserved by local operations. The different sectors can be distinguished by attaching to each sector a label, which we refer to as the sector's ``charge.'' Therefore, an equivalent way to characterize a superselection rule is to say that the charge is locally conserved. In the context of a cryptographic protocol, this means that when one of the parties (Alice, say) performs an operation, the charge in Alice's laboratory is preserved.

An important special case arises if the Hilbert space ${\cal H}$ transforms as a unitary representation of a compact group $G$, and the sectors are labeled by the irreducible representations of $G$. An equivalent way to describe the superselection rule in that case is to say that the allowed operations must commute with the action of $G$ on ${\cal H}$. In fact, it has been shown by Doplicher and Roberts \cite{doplicher} that such superselection rules are almost the most general ones allowed under rather weak conditions that apply in particular to quantum field theories (without gravity) in three or more spatial dimensions. We say ``almost'' because there is an additional freedom to assign to a localized state an even or odd fermion number. This fermion number is more than just a conserved charge, because of the property that the wave function changes sign when two fermions are exchanged.

In two spatial dimensions, there is a richer classification of superselection rules, reflecting the exotic quantum numbers carried by pointlike nonabelian anyons that occur in topological quantum field theories \cite{rehren,kitaev_anyon,freedman}. We will postpone further discussion of nonabelian anyons until Sec.~\ref{sec:two-party}, concentrating for now on the superselection rules associated with compact symmetry groups (and ignoring fermions).

An important example is the group $U(1)$ associated with conservation of the electric charge $Q$. An agent acting locally can create or annihilate pairs of particles that carry equal and opposite charges, but cannot change the total charge in her vicinity. In particular, this agent is unable to transform any eigenstate of $Q$ into a coherent superposition of states with different charges, as emphasized by Wick, Wightman, and Wigner \cite{wick52,wick70}.

While we might readily accept that local creation of electric charge is physically impossible, other conservation laws impose superselection rules that do more violence to our intuition. Suppose, for example (in nonrelativistic quantum mechanics), that our agent's actions are required to conserve the angular momentum $\vec J$ locally. Are we to conclude that if the agent is presented with a spin-$1/2$ object polarized spin-up along the $z$ axis, it is impossible for him to transform it to a coherent superposition of the spin-up and spin-down states? How are we to describe what happens when a magnetic field is turned on pointing in the $x$ direction and the spin begins to precess? A partial resolution of this puzzle is attained by noting that the angular momentum of a classical magnet has an uncertainty large compared to $\hbar$, so that conservation of angular momentum need not prevent the magnet from coherently exchanging $J_z=\hbar$ with the spin. But this explanation does not fully address how the existence of the classical magnet is itself compatible with the superselection rule.

Such issues were cogently discussed many years ago by Aharonov and Susskind \cite{aharonov}. They emphasized that even if the total angular momentum has a definite value (like zero), we can still speak sensibly of the {\em relative} orientation of two subsystems. Whenever an experimentalist observes the precession of a spin, it is implicit that a reference state has been established that in effect breaks the rotational symmetry, and that the precession is measured relative to this reference standard. Furthermore, Aharonov and Susskind \cite{aharonov} emphasized that just as conservation of angular momentum need not prevent us from measuring the relative angular orientation of two objects, so the charge superselection rule need not prevent us from measuring relative phases in superpositions of states of different charge. 

\subsection{Abelian case}
Before we discuss the more general case in which the symmetry may be nonabelian, it will be useful to consider the symmetry group $G=U(1)$. 
Then the charge operator $Q$ (the generator of $G$) has eigenvalues $q\in Z$, and we denote the corresponding orthonormal eigenstates by $|q\rangle$. Formal states of definite phase (with continuum normalization) can be constructed as
\begin{equation}
|\theta\rangle = {1\over \sqrt{2\pi}} \sum_{q=-\infty}^{\infty}e^{-iq\theta}|q\rangle \quad (0\le \theta < 2\pi)~,
\end{equation}
where
\begin{equation}
\langle \theta'|\theta\rangle = {1\over 2\pi}\sum_{q=-\infty}^\infty e^{-iq(\theta-\theta')}=\delta(\theta'-\theta)~,
\end{equation}
and 
\begin{equation}
|q\rangle = {1\over \sqrt{2\pi}}\int_0^{2\pi} d\theta~e^{iq\theta}|\theta\rangle~.
\end{equation}
The phase state $|\theta\rangle$ is the improper eigenstate with eigenvalue $e^{i\theta}$ of the unitary operator
\begin{equation}
U_+=\sum_{q=-\infty}^\infty |q+1\rangle\langle q|
\end{equation}
that increments the value of the charge by one unit. While the phase $\theta$ is physically unobservable due to the charge superselection rule, the relative phase of $\theta'-\theta$ of the two states $|\theta'\rangle$ and $|\theta\rangle$ commutes with the charge operator $Q$ and so is measurable in principle. Indeed, the state 
\begin{eqnarray}
\int_0^{2\pi}d\theta''~|\theta' +\theta''\rangle\otimes |\theta+\theta''\rangle\nonumber\\
=\sum_{q=-\infty}^\infty e^{-iq(\theta-\theta')}|-q\rangle\otimes |q\rangle
\label{charge_ent}
\end{eqnarray}
has a definite value of the relative phase $\theta'-\theta$ and total charge zero. That is, it is an (unnormalizable) eigenstate with eigenvalue $e^{i(\theta-\theta')}$ of the charge-conserving operator $U_-\otimes U_+$, where $U_-=U_+^\dagger$.  

Similarly, the phases $\phi_q$ appearing in the expansion of the state $|\psi\rangle_A$ of a system $A$, 
\begin{equation}
|\psi\rangle_A =\sum_q \psi_q e^{-iq\phi_q}|q\rangle_A
\end{equation}
(where the $\psi_q$'s are real and positive), are themselves unobservable, but they can be meaningfully compared to the phases appearing in the state $|\theta\rangle_R$ of a charge reservoir $R$. 
For example, by projecting $|\theta\rangle_R\otimes |\psi\rangle_A$ onto the sector with total charge zero we obtain the state
\begin{eqnarray}
&&|\psi\rangle_{RA}={1\over\sqrt{2\pi}}\int d\theta'~ |\theta+\theta'\rangle_R\otimes e^{-iQ\theta'}|\psi\rangle_A \nonumber\\
&&=\sum_{q} \psi_qe^{-iq(\phi_q-\theta)}|-q\rangle_R \otimes |q\rangle_A
\end{eqnarray}
which has measurable relative phases. A state like $|\theta\rangle_R$ of a charge reservoir $R$ that provides a phase standard with which other states can be compared will be called a ``reference state'' or a ``condensate.''

In the state $|\psi\rangle_{RA}$, the charge of the system $A$ is compensated (``screened'') by the charge of the reservoir $R$. Therefore, the system and reservoir are entangled, and tracing out the reservoir destroys the coherence of the superposition of charge states for the system. While formally correct, this statement can be misleading if the reservoir remains accessible and is allowed to interact with the system during subsequent operations. For example, the operator $\left(U_+\right)_A$ that increases the charge of the system by one unit is disallowed by the superselection rule, but it can be accurately simulated by the allowed charge-conserving operator $\left(U_-\right)_R\otimes\left(U_+\right)_A$ acting on $|\psi\rangle_{RA}$  --- this operator increases the charge of $A$ by borrowing a unit of charge from $R$. If the reservoir remains accessible at all times, then an arbitrary (not necessarily charge conserving) operation acting on $A$ can be perfectly simulated by a charge-conserving operation acting on $RA$. Thus, at least as a matter of principle, the charge superselection rule places no inescapable restrictions on the allowed operations. This is the main point stressed by Aharonov and Susskind \cite{aharonov}.

The phase reference state can be interpreted physically as a static piece of superconducting material with a definite value of the superconducting phase. While the phase itself is not gauge-invariant, the relative phase of the system and reservoir  has observable consequences (like the Josephson effect) when the two are brought into contact. Similar issues, discussed in \cite{molmer,rudolf,fuchs,bartlett,vanenk}, arise when considering the physical content of relative phases in optical systems.

\subsection{Nonabelian case}
\label{subsec:nonabelian}

Our discussion of the abelian case has suggested that superselection rules are nullified if suitable reference systems are available. Now we consider the more general case, where the symmetry group is $G$, which may be either a finite group or a compact Lie group. The superselection rule dictates that allowed local operations must commute with $G$. But we may anticipate that if a condensate is accessible that completely breaks the $G$ symmetry, then in effect there is no operative symmetry at all, and the superselection rules place no restrictions on the allowed operations. 

Formally, if the symmetry is completely broken, then the possible orientations of the condensate are in one-to-one correspondence with the elements of the symmetry group $G$. In a particular ``fixed gauge,'' the states of the condensate are denoted $|\phi\rangle$ where $\phi\in G$, and these states transform as the left regular representation of $G$. That is, a symmetry transformation $g\in G$ acting on the condensate is represented by the unitary $U(g)$  where
\begin{equation}
U(g)|\phi\rangle \to |g\phi\rangle~.
\end{equation}
These states can be expanded in the basis of irreducible representations of $G$ as
\begin{equation}
|\phi\rangle = \sum_{q,i,a}\sqrt{n_q\over n_G}~D^q_{ia}(\phi)|q,i,a\rangle~,
\label{phi_expand}
\end{equation}
where $n_q$ denotes the dimension of the irreducible representation $D^q(\phi)$ and $n_G$ is the order of $G$. Inverting the Fourier transform we obtain
\begin{equation}
|q,i,a\rangle= \sum_{\phi\in G}\sqrt{n_q\over n_G}~D^{q*}_{ia}(\phi)|\phi\rangle~.
\label{q_expand}
\end{equation}
Note that in eq.~(\ref{phi_expand},\ref{q_expand}) we have used notation appropriate for a finite group; in the case of a compact Lie group, the sum over $\phi\in G$ would be replaced by an integral with respect to an invariant measure on the group. The states $|q,i,a\rangle$ transform under $G$ as
\begin{equation}
\label{gauge_transform}
U(g)|q,i,a\rangle = \sum_j |q,j,a\rangle D^q_{ji}(g)~.
\end{equation}

In keeping with standard physics terminology, we will refer to the index  $i=1,2,\dots,n_q$ in $|q,i,a\rangle$ as the ``color index,'' and to the action eq.~(\ref{gauge_transform}) of $U(g)$ on this index as a ``gauge transformation.'' The index $a=1,2,\dots,n_q$, distinguishing the $n_q$ copies of the representation $D^q$ that occur in the decomposition of the regular representation, will be called the ``flavor'' index. The physical ``$G$-invariant'' operations are those that commute with all gauge transformations --- these preserve $q$ and act nontrivially only on the flavor, not the color. Therefore, by including the color we have chosen a redundant description of the physical Hilbert space. This redundancy, while not absolutely necessary, is quite convenient, and in particular will be useful for our discussion in Sec.~\ref{sec:simulation} of the security of quantum protocols.

In addition to the $G$ gauge symmetry, there is also a group $G$ of ``global'' transformations that commute with $U(g)$, under which the states $|\phi\rangle$ transform as the right regular representation of $G$; the element $h$ of the global group is represented by $V(h)$ where  
\begin{equation}
V(h)|\phi\rangle = |\phi h^{-1}\rangle~,
\end{equation}
and
\begin{equation}
V(h)|q,i,a\rangle = \sum_b|q,i,b\rangle D^{q*}_{ba}(h)~.
\end{equation}
Thus the global transformations act on the flavor index $a$ of the states in the  $\{|q,i,a\rangle\}$ basis --- unlike the gauge transformations, they act nontrivially on the physical states.

In more geometric terms, a condensate may be interpreted as an asymmetric classical rigid body that can be rotated either ``actively'' or ``passively.'' What we have called the color (gauge) rotation is a passive rotation that acts on the space-fixed axes --- it does not change the actual orientation of the body but only changes our mathematical description of the orientation. In contrast, what we have called the flavor (global) rotation is an active rotation that acts on the body-fixed axes and alters the physical orientation. A flavor rotation is $G$-invariant in the sense that it commutes with color rotations, and so is a physical operation, allowed by the superselection rule. 

In contrast to the flavor orientation, the color orientation of an isolated system $A$ has no invariant meaning, as it is modified by a color rotation. However, the orientation of $A$ {\em relative to the condensate $R$} does have meaning, and an operator that rotates the relative orientation admits an invariant description. Suppose, for example, that system $A$ is itself a condensate in the state $\phi_A$, while the state of $R$ is $\phi_R$. The relative orientation 
\begin{equation}
\phi_{\bar R A}\equiv \phi_R^{-1} \phi_A
\end{equation}
is invariant if a common color rotation 
\begin{equation}
U(h)_{RA}: \phi_A\to h\phi_A~, \quad \phi_R\to h\phi_R
\end{equation}
is applied to both objects. 
The transformation $U(g)^{\rm inv}_{RA}$ that changes the relative orientation according to
\begin{equation}
U(g)^{\rm inv}_{RA}: \phi_{\bar RA} \to g\phi_{\bar RA}~,
\end{equation}
has an invariant meaning and commutes with the color rotation $U(h)_{RA}$. 
We may interpret the invariant rotation as one that rotates $A$ while $R$ is ``held fixed,'' acting as
\begin{equation}
U(g)^{\rm inv}_{RA}~\big(|\phi_R\rangle\otimes |\phi_A\rangle \big)= |\phi_R\rangle\otimes |\phi_R g\phi_R^{-1}\phi_A\rangle~,
\end{equation}
or equivalently
\begin{equation}
U(g)^{\rm inv}_{RA}= \sum_{\phi\in G}\left(|\phi\rangle\langle \phi|\right)_R\otimes U(\phi g\phi^{-1})_A~.
\end{equation}

If system $A$ is not a reference system but rather an object transforming as the irreducible representation $q$ of $G$, then $U(\phi g \phi^{-1})$ can be expanded as
\begin{eqnarray}
&&U(g)^{\rm inv}_{RA}= \sum_{\phi\in G}\left(|\phi\rangle\langle \phi|\right)_R\nonumber\\
&&\otimes \left(\sum_{i,j,a,b} |q,i\rangle D^q_{ia}(\phi)D^q_{ab}(g)D^q_{bj}(\phi^{-1})\langle q,j|\right)_A.
\end{eqnarray}
More generally, any transformation
\begin{equation}
M_A: |q,i\rangle_A \to \sum_j|q,j\rangle_A M_{ji}
\end{equation}
acting on the color degree of freedom can be simulated by the invariant operation
\begin{eqnarray}
&&M^{\rm inv}_{RA}= \sum_{\phi\in G}\left(|\phi\rangle\langle \phi|\right)_R\nonumber\\
&&\otimes \left(\sum_{i,j,a,b} |q,i\rangle D^q_{ia}(\phi)M_{ab}D^q_{bj}(\phi^{-1})\langle q,j|\right)_A.
\label{inv_op_sumover_group}
\end{eqnarray}
$M^{\rm inv}_{RA}$ has an invariant meaning because it transforms the color of $A$ relative to the color of the reference system $R$; in effect, the color rotation is simulated by converting the color index into a flavor index (depending on $\phi$), on which $M$ may act with impunity. For fixed $\phi$, the simulation is achieved via the isomorphism
\begin{equation}
|q,a\rangle_A\to |q,\phi,a\rangle_{RA} \equiv|\phi\rangle_R\otimes \sum_j |q,j\rangle_A D^q_{ja}(\phi)~,
\label{invariant_basis}
\end{equation}
such that
\begin{equation}
M^{\rm inv}_{RA}~|q,\phi,a\rangle_{RA}=\sum_b|q,\phi,b\rangle_{RA} M_{ba}~.
\label{invariant_reference}
\end{equation}
Furthermore, this isomorphism can  be extended to operators $M$ that change the value of $q$ as well as rotating the color for fixed $q$;  the operator
\begin{equation}
M_A: |q,i\rangle_A \to \sum_{q',j}|q',j\rangle_A M^{q'q}_{ji}~,
\end{equation}
is simulated by 
\begin{equation}
M^{\rm inv}_{RA}~|q,\phi,a\rangle_{RA}=\sum_{q',b}|q',\phi,b\rangle_{RA} M^{q'q}_{ba}~,
\label{any_transform}
\end{equation}
which generalizes the result
\begin{equation}
M^{\rm inv}_{RA}~\left(|\theta\rangle_R\otimes e^{-iq\theta}|q\rangle_A\right)=|\theta\rangle_R\otimes \sum_{q'}e^{-iq'\theta}|q'\rangle M^{q'q}
\end{equation}
that we found in the case of $G=U(1)$.

\subsection{Properties of the simulation}
\label{subsec:properties}

We will refer to the world in which all operations are required to commute with the action of the symmetry group $G$ as the ``invariant world'' or ``$I$-world,'' and we refer to the world in which arbitrary operations are allowed as the ``unrestricted world'' or ``$U$-world.'' What we have observed in eq.~(\ref{invariant_basis},\ref{any_transform}) is that the physics of the $U$-world can be faithfully reproduced in the $I$-world, as long as a suitable reference system is at our disposal.

Let us restate the main conclusion in a more succinct notation: Suppose $A$ is an arbitrary system that transforms as some representation of the group $G$, and let $R$ be a ``reference system'' that transforms as the left regular representation of $G$. Let $M$ be an arbitrary transformation acting on $A$. Then there is a corresponding transformation $M^{\rm inv}$ acting on $R$ and $A$ defined as
\begin{eqnarray}
\label{invariant_succinct}
M^{\rm inv}=\sum_{\phi\in G}\big(|\phi\rangle\langle\phi|\big)_R \otimes \big(U(\phi)MU(\phi)^{-1}\big)_A~.
\end{eqnarray}
$M^{\rm inv}$ is an invariant operator whose action on $RA$ {\em simulates} the action of $M$ on $A$.

That is, the operators $M^{\rm inv}$ have the following easily verified properties:

\begin{enumerate} 
\item{\em $M^{\rm inv}$ is $G$-invariant}. 

Proof: From the transformation properties of $R$ and $A$ we have
\begin{eqnarray}
&&\big(U(g)\otimes U(g)\big)M^{\rm inv}\big(U(g)^{-1}\otimes U(g)^{-1}\big)\nonumber\\
&&= \sum_{\phi\in G}\big(|g\phi\rangle\langle g\phi|\big) \otimes \big(U(g\phi)MU(g\phi)^{-1}\big)\nonumber\\
&& = M^{\rm inv}~,
\end{eqnarray}
where in the last step we have reparametrized the sum by replacing $\phi\to g^{-1}\phi$.

\item {\em Invariant operators on $RA$ provide a representation of operators on $A$}. 

Proof: We have 
\begin{eqnarray}
&&M_1^{\rm inv}M_2^{\rm inv}= \sum_{\phi_1,\phi_2\in G}\big( |\phi_1\rangle\langle \phi_1|\phi_2\rangle\langle \phi_2|\big)\nonumber\\
&&\quad \otimes \big(U(\phi_1)M_1U(\phi_1^{-1})U(\phi_2)M_2 U(\phi_2^{-1})\big)\nonumber\\
&&= \sum_{\phi\in G}\big(|\phi\rangle\langle\phi|\big)\otimes \big(U(\phi)M_1M_2U(\phi)^{-1}\big)\nonumber\\
&& =\big(M_1M_2\big)^{\rm inv}~.
\end{eqnarray}

\item {\em If $M$ is $G$-invariant, then $M^{\rm inv}= I_R\otimes M_A$.} 

Proof: If $U(\phi)$ commutes with $M$ for each $\phi$, then
\begin{eqnarray}
M^{\rm inv}= \sum_{\phi\in G}\big(|\phi\rangle\langle\phi|\big)\otimes M\big)= I\otimes M~.
\end{eqnarray}
\item {\em If $\rho$ is invariant and ${\rm tr}(\rho_R)=1$, then 
\begin{equation}
\label{prob_same}
{\rm tr}~M^{\rm inv}\big(\rho_R\otimes \rho\big)={\rm tr}~ M \rho ~.
\end{equation}
}
Proof: If $U(\phi)$ commutes with $\rho$ for each $\phi$, then
\begin{eqnarray}
&&{\rm tr}~M^{\rm inv}\big(\rho_R\otimes \rho\big)\nonumber\\
&&=\sum_{\phi\in G}\langle\phi|\rho_R|\phi\rangle~{\rm tr}\big(M ~U(\phi)^{-1}\rho U(\phi)\big)\nonumber\\
&&={\rm tr} \big(\rho_R \big)\cdot {\rm tr} \big(M \rho \big)={\rm tr}~\big( M \rho \big)~.
\end{eqnarray}
\end{enumerate}

The properties 1 and 4 mean that as long as the state $\rho$ of $A$ is $G$-invariant, then by making use of a reference system, measurements in the $U$-world can be faithfully simulated by measurements in the $I$-world. That is, given an arbitrary measurement performed on $A$ (with operation elements that are not necessarily $G$-invariant), there is an invariant measurement performed on $RA$ (with $G$-invariant operation elements) that has the same probability distribution of outcomes. Furthermore, it follows from property 2 that the physics of the $U$-world can be faithfully reproduced in the $I$-world even if the measurement is preceded by a series of unitary transformations --- applying $V^{\rm inv}$ in the $I$-world has the same effect as applying $V$ in the $U$-world. Property 3 tells us that, as expected, the reference system $R$ is superfluous if the $U$-world transformation acting on $A$ is already $G$-invariant.

To derive these properties, we require that the reference system transform as the regular representation of $G$, but no condition is needed on the {\em state} $\rho_R$ of the reference system. Loosely speaking, the reference system is needed so that when a noninvariant operation acts on $A$, the change in the charge of $A$ can be balanced by a compensating change in the charge of $R$. But if the state $\rho$ of $A$ is invariant, then only the charge-conserving part of $M$ contributes to the expectation value ${\rm tr}(M\rho)$ anyway. In the simulation of this charge-conserving part of $M$, the reference system is superfluous and its state irrelevant.

Note that if $G$ is a Lie group rather than a finite group, then the regular representation is infinite dimensional, and our formal arguments require $R$ to be an infinite-dimensional system. How is the fidelity of the simulation affected if $R$ is truncated to a finite-dimensional system? In fact, the fidelity will still be perfect if the charge remains bounded in the process to be simulated. Consider, for example, the case $G=U(1)$, for which eq.~(\ref{invariant_succinct}) becomes, e.g., 
\begin{equation}
\big(|q-r\rangle \langle q|\big)^{\rm inv}= \sum_{q'}\big(|q'+r\rangle\langle q'|\big)_R\otimes \big(|q-r\rangle \langle q|\big)_A~;
\end{equation}
in the $I$-world, a process in which $r$ units of charge are removed from $A$ is simulated by adding the $r$ units to $R$. Suppose we are assured that the total charge added to or removed from $A$ will never exceed $r$ units. Then we may choose the initial state of $R$ to carry charge zero, and we can limit $R$ to the $2r+1$ dimensional space spanned by the states $|q_R\rangle,~q_R=-r, -r+1, \dots r-1, r$. This truncated reference system suffices because states with $|q_R|> r$ will never be accessed in the simulation anyway. A similar remark applies if $G$ is an arbitrary compact Lie group.

\section{Reference systems and quantum protocols}
\label{sec:simulation}

We have concluded that in the presence of a suitable reference system, superselection rules place no inescapable restrictions on the allowed operations. We may anticipate, therefore, that a cryptographic protocol is secure in the invariant ``$I$-world'' (governed by the superselection rule) if and only if it is secure in the unrestricted ``$U$-world.'' If we faithfully adhere to the usual stringent principles of quantum cryptology and place no restrictions on the resources available to our adversaries, then we must admit the possibility that the dishonest parties could share access to a reference system during the execution of the protocol. For the case of superselection rules arising from compact symmetry groups, this observation suffices to answer Popescu's question about the impact of superselection rules on the security of quantum protocols.

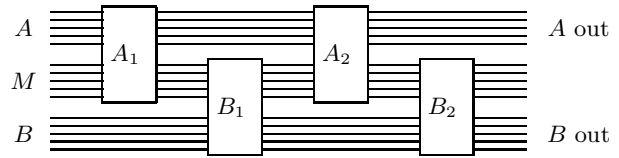
\begin{figure}
\centering
\begin{picture}(220,60)
\put(0,40){\makebox(0,12){$A$}}
\put(0,20){\makebox(0,12){$M$}}
\put(0,0){\makebox(0,12){$B$}}

\put(200,40){\makebox(20,12){$A$ out}}
\put(200,0){\makebox(20,12){$B$ out}}

\put(10,0){\line(1,0){60}}
\put(10,3){\line(1,0){60}}
\put(10,6){\line(1,0){60}}
\put(10,9){\line(1,0){60}}
\put(10,12){\line(1,0){60}}

\put(10,20){\line(1,0){20}}
\put(10,23){\line(1,0){20}}
\put(10,26){\line(1,0){20}}
\put(10,29){\line(1,0){20}}
\put(10,32){\line(1,0){20}}

\put(10,40){\line(1,0){20}}
\put(10,43){\line(1,0){20}}
\put(10,46){\line(1,0){20}}
\put(10,49){\line(1,0){20}}
\put(10,52){\line(1,0){20}}

\put(30,18){\framebox(20,36){$A_1$ }}
\put(70,-2){\framebox(20,36){$B_1$ }}

\put(50,20){\line(1,0){20}}
\put(50,23){\line(1,0){20}}
\put(50,26){\line(1,0){20}}
\put(50,29){\line(1,0){20}}
\put(50,32){\line(1,0){20}}

\put(50,40){\line(1,0){60}}
\put(50,43){\line(1,0){60}}
\put(50,46){\line(1,0){60}}
\put(50,49){\line(1,0){60}}
\put(50,52){\line(1,0){60}}

\put(90,20){\line(1,0){20}}
\put(90,23){\line(1,0){20}}
\put(90,26){\line(1,0){20}}
\put(90,29){\line(1,0){20}}
\put(90,32){\line(1,0){20}}

\put(90,0){\line(1,0){60}}
\put(90,3){\line(1,0){60}}
\put(90,6){\line(1,0){60}}
\put(90,9){\line(1,0){60}}
\put(90,12){\line(1,0){60}}

\put(110,18){\framebox(20,36){$A_2$ }}
\put(150,-2){\framebox(20,36){$B_2$ }}

\put(170,20){\line(1,0){20}}
\put(170,23){\line(1,0){20}}
\put(170,26){\line(1,0){20}}
\put(170,29){\line(1,0){20}}
\put(170,32){\line(1,0){20}}

\put(130,40){\line(1,0){60}}
\put(130,43){\line(1,0){60}}
\put(130,46){\line(1,0){60}}
\put(130,49){\line(1,0){60}}
\put(130,52){\line(1,0){60}}

\put(130,20){\line(1,0){20}}
\put(130,23){\line(1,0){20}}
\put(130,26){\line(1,0){20}}
\put(130,29){\line(1,0){20}}
\put(130,32){\line(1,0){20}}

\put(170,0){\line(1,0){20}}
\put(170,3){\line(1,0){20}}
\put(170,6){\line(1,0){20}}
\put(170,9){\line(1,0){20}}
\put(170,12){\line(1,0){20}}

\end{picture}
\caption{A two-player quantum game. Alice and Bob have private systems, and a message system that they pass back and forth. At the end of the game, Alice and Bob measure their private systems.}
\label{fig:two-player}
\end{figure}

Let us now discuss this point in greater detail. To be explicit, consider at first a protocol involving two parties, Alice and Bob. Alice holds a private local system $A$ that is beyond Bob's control, and Bob holds a private local system $B$ that is beyond Alice's control. In addition, there is a message system $M$ that they can pass back and forth. At the beginning of the protocol, they share a product state $\rho_A\otimes\rho_B\otimes \rho_M$. In each round of the protocol, one of the parties performs a  joint quantum operation on her/his local system and the message, and then sends the message system to the other party. Finally, after all quantum communication is completed, both parties perform local measurements. (See Fig.~\ref{fig:two-player}.)

For example, the goal of the protocol might be to flip an unbiased coin. In that case, the final measurement performed by each party has two possible outcomes, 0 or 1. If both parties follow the protocol, then both obtain the same outcome. Furthermore, the two outcomes are equiprobable. A coin flipping protocol is {\em secure} if neither party, by departing from the protocol, can bias significantly the outcome of the other party's measurement. 

We say that a {\em strong} coin flipping protocol has bias $\varepsilon$ if neither party by cheating can force {\em either} outcome to occur with probability greater than ${1\over 2}+\epsilon$. In a {\em weak} coin flipping protocol, Alice wins if the outcome is 0 and Bob wins if the outcome is 1, and we say that the bias is $\varepsilon$ if neither can force a {\em win} with probability greater than ${1\over 2}+\epsilon$. (Thus, in a weak protocol with bias $\varepsilon$, a cheater might be able to {\em lose} on purpose with a probability exceeding ${1\over 2}+\epsilon$). Note that the protocol might abort if cheating is detected; by ``the probability of outcome 0'' we mean the joint probability that the protocol does not abort and the outcome is 0. Kitaev \cite{kitaev_coin,ambainis_kitaev} has shown that, if no superselection rules are imposed, then strong quantum coin flipping is impossible with bias $\varepsilon < {1\over\sqrt{2}}-{1\over 2}=.207$. Ambainis \cite{ambainis_coin} has shown that a weak coin flipping protocol with bias $\varepsilon$ requires at least $\Omega(\log\log {1\over\varepsilon})$ rounds of communication. 

We are interested in whether these conclusions about coin flipping in the $U$-world remain valid in the $I$-world. For a coin flipping protocol in the $I$-world, we may assume that the initial state shared by Alice and Bob is a tensor product of invariant states $\rho_A\otimes \rho_B\otimes \rho_M$. In the honest protocol, Alice and Bob take turns applying $G$-invariant operations to the system that they share, then measure invariant observables. In fact, without loss of generality, we may assume \cite{mayers_qbc} that each operation applied by Alice or Bob is an invariant unitary transformation, and that the final measurement is an invariant projective measurement.

If Alice and Bob play the game honestly, then the probability $P_B(b)$ that Bob's measurement yields the particular outcome $b$ can be expressed as
\begin{equation}
P_B(b)={\rm tr} \Big(E_{B,b}~ V\big(\rho_A\otimes\rho_B\otimes \rho_M\big)V^\dagger\Big)~,
\end{equation}
where\begin{equation}
V= V_{B_n}V_{A_n} \dots V_{B_2}V_{A_2}V_{B_1}V_{A_1}~.
\end{equation}
Here the $V_{A_j}$ are unitary transformations applied to $AM$ (we have assumed that Alice makes the first move in the game), the $V_{B_j}$ are unitary transformations applied to $BM$, and the $E_{B,b}$ are the projectors defining Bob's final measurement. Furthermore, in the $I$-world protocol,  $V_{A_j}$, $V_{B_j}$, and $E_{B,b}$ are $G$-invariant. In effect, then, Bob measures the invariant operator
\begin{equation}
F_{B,b}=V^\dagger E_{B,b} V
\end{equation}
in the invariant state $\rho_A\otimes\rho_B\otimes \rho_M$.

Of course, a protocol in the $I$-world can be regarded as a special case of a protocol in the $U$-world, where the initial state is a product state, and Kitaev's result applies to this $U$-world protocol. Therefore, one of the parties (Alice, say) can force one of the outcomes (0, say) with probability at least ${1\over\sqrt{2}}$. However, Alice's cheating strategy that achieves this result might employ operations that are not $G$-invariant. To show that Kitaev's result also applies to the original $I$-world protocol, we must show that Alice's cheating strategy in the $U$-world can be faithfully simulated in the $I$-world by making use of a suitable reference system. For this purpose, we apply the properties of the invariant operator $M^{\rm inv}$ that were discussed in Sec.~\ref{subsec:properties}.

When Alice cheats in the $U$-world, she replaces the operator $V_{A_j}$ called for in the honest protocol with an arbitrary operator $V_{A_j}'$ applied to $AM$, where $V_{A_j}'$ is not necessarily $G$-invariant.  Then Bob's measurement yields the outcome $b$ with probability 
\begin{equation}
P_B'(b)= {\rm tr} \Big(F_{B,b}'\big(\rho_A\otimes\rho_B\otimes \rho_M\big)\Big)~,
\end{equation}
where
\begin{equation}
F_{B,b}'= V'^{\dagger}E_{B,b} V'
\end{equation}
and
\begin{equation}
V'= V_{B_n}V_{A_n}' \dots V_{B_2}V_{A_2}'V_{B_1}V_{A_1}'~.
\end{equation}

This cheating strategy in the $U$-world can be simulated in the $I$-world if Alice has a reference system $R$ --- instead of applying the noninvariant operator $V_{A_j}'$ to the system $AM$, she applies the invariant operator $V_{A_j}'^{\rm inv}$ to $RAM$. Note that since Bob follows the honest protocol, which requires $V_{B_j}$ to be $G$-invariant, applying $V_{B_j}$ to $BM$ is equivalent to applying $V_{B_j}^{\rm inv}$ to $RBM$, by property 3 in Sec.~\ref{subsec:properties}. Therefore, when Alice adopts the $I$-world strategy, Bob obtains outcome $b$ with probability
\begin{equation}
\simu{P}_B'(b)= {\rm tr} \Big(\simu{F}_{B,b}'\big(\rho_R\otimes\rho_A\otimes\rho_B\otimes \rho_M\big)\Big)
\end{equation}
where 
\begin{equation}
\simu{F}_{B,b}'= \simu{V}'^{\dagger}E_{B,b} \simu{V}'
\end{equation}
and
\begin{equation}
\simu{V}'= V_{B_n}^{\rm inv}V_{A_n}'^{\rm inv} \dots V_{B_2}^{\rm inv}V_{A_2}'^{\rm inv}V_{B_1}^{\rm inv}V_{A_1}'^{\rm inv}~.
\end{equation}
But since the invariant operators provide a representation (property 2), we may write $\simu{V}'=V'^{\rm inv}$, and since $E_{B,b}=E_{B,b}^{\rm inv}$ as well, we have
\begin{equation}
\simu{F}_{B,b}'= F_{B,b}'^{\rm inv}~.
\end{equation}
Finally, the initial state $\rho_A\otimes\rho_B\otimes \rho_M$ shared by Alice and Bob is $G$-invariant; therefore, by property 4, 
\begin{equation}
\simu{P}_{B}'(b)= {P}_{B}'(b)~;
\end{equation}
the measurement outcome $b$ in the $I$-world protocol occurs with the same probability as the outcome $b$ in the $U$-world protocol.

Therefore, Alice's simulated cheating strategy in the $I$-world perfectly reproduces the probability distribution for Bob's measurement outcome that is achieved by her cheating strategy in the $U$-world. The same is true if Bob makes the first move in the game instead of Alice. Similarly, if Bob is the cheater, Bob has a strategy in the $I$-world that simulates his $U$-world cheating strategy. We conclude that if Alice (or Bob) can cheat in the $U$-world, then she (he) can cheat just as successfully in the $I$-world. Thus, Kitaev's proof of the impossibility of strong coin flipping with bias $\epsilon < {1\over\sqrt{2}} - {1\over 2}$, originally formulated in the $U$-world, also applies to the $I$-world. Similarly, Ambainis's lower bound on the number of rounds of communication needed for weak coin flipping also applies to the $I$-world.

This conclusion that cheating in the $U$-world can be successfully simulated in the $I$-world applies not just to coin flipping protocols, but to any two-party protocol in which the goal of a cheating Alice is to bias the outcome of a measurement performed by an honest Bob. Furthermore, it is straightforward to generalize the argument to an $n$-party protocol, in which $k$ cheating parties wish to bias the outcomes of measurements performed by the $n-k$ honest parties. For such a protocol in the $I$-world, where the initial state is a product of invariant states, any cheating strategy that can be executed in the $U$-world can be simulated perfectly in the $I$-world if the $k$ cheating parties share access to a reference system. Therefore, the protocol can be no more secure in the $I$-world than in the $U$-world.

To summarize: Let us refer to an $n$-party quantum game as an $I$-world game if the initial state is a product of invariant states, and  if in the honest protocol all operations performed by the parties are invariant operations. If $k < n$ parties are cheaters, we say that their cheating strategy is an $I$-world cheating strategy if the cheaters are required to perform invariant operations, and we say that their cheating strategy is a $U$-world cheating strategy if the operations performed by the cheaters are unrestricted. Let us say that an $I$-world cheating strategy is {\em equivalent} to a $U$-world cheating strategy if both strategies produce the same probability distributions for the outcomes of the measurements performed by the $n-k$ honest parties. 
We have proved:

\medskip
\noindent{\bf Theorem 1} {\em Suppose that in the $I$-world all quantum operations are required to be $G$-invariant, where $G$ is a compact Lie group, and that in the $U$-world quantum operations are unrestricted. Consider an $n$-party $I$-world quantum game, and a $U$-world cheating strategy ${A}'$ in which $k< n$ parties cheat. Then there is an $I$-world cheating strategy $\simu{A}'$ that is equivalent to ${A}'$.}
\medskip

\noindent 
As we observed in Sec.~\ref{subsec:properties}, the reference system required by the cheaters in the $I$ world can be finite-dimensional, as long as the cheaters in the $U$-world apply operations that change the ``charge'' by a bounded amount.

\section{Distributed reference systems}
\label{sec:distributed}
The key ingredient in our discussion of $I$-world quantum protocols is the observation that $G$-noninvariant operations can be faithfully simulated through the use of a reference system. Suppose, for example, that Alice and Bob take turns acting on a system $C$ that they pass back and forth. Then Alice and Bob in the $I$-world can simulate an arbitrary $U$-world protocol in which the initial state of $C$ is $G$-invariant. They carry out the simulation by passing the reference system $R$ back and forth along with $C$, each taking turns applying invariant operations to $RC$. Similarly, in our analysis of cheating in Sec.~\ref{sec:simulation}, we allowed the $k$ cheaters to pass the reference system $R$ among themselves as needed during the execution of the protocol. A reference system that travels from place to place might be called {\em itinerant}.

Here we will briefly discuss an alternative scenario, in which the parties share a {\em distributed} reference system --- each party holds a fixed portion of this system throughout the execution of the protocol. This discussion is not actually needed for our analysis of security, but it is helpful nonetheless for understanding the physics of superselection rules. Indeed, in many physical situations in which reference systems are used (e.g., in optical physics), the system is distributed rather than itinerant.

Let $A$ denote Alice's part of the reference system, $B$ denote Bob's part, and suppose that at the start of the protocol $AB$ is prepared in the state
\begin{equation}
\label{0_AB}
|0\rangle_{AB}= {1\over \sqrt{n_G}}\sum_{\phi\in G}|\phi\rangle_A\otimes |\phi\rangle_B~.
\end{equation}
This state has trivial total charge; indeed, when expressed in the Fourier-transformed charge-eigenstate basis, it is
\begin{equation}
|0\rangle_{AB}= {1\over\sqrt{n_G}}\sum_{q,i,a}|\bar q,i,a\rangle_A\otimes |q,i,a\rangle_B~.
\end{equation}
Thus, in principle Alice (say) could prepare $|0\rangle_{AB}$ in her lab and then ship half of it to Bob. (The state $|0\rangle_{AB}$ is unnormalizable and unphysical if $G$ is a Lie group. For now we will suppose that $G$ is a finite group, but we will comment on the case of a Lie group below.)

In the state $|0\rangle_{AB}$, Alice's condensate, and Bob's, have values that are distributed uniformly over the group $G$, but these values are locked together. Therefore, if $|\psi\rangle_C$ is any pure state of $C$, then $M^{\rm inv}_{AC}$ and $M^{\rm inv}_{BC}$ act on $|0\rangle_{AB}\otimes|\psi\rangle_C $ in the same way:
\begin{eqnarray}
&&M^{\rm inv}_{AC}\big(|0\rangle_{AB}\otimes |\psi\rangle_C\big)=M^{\rm inv}_{BC}\big(|0\rangle_{AB}\otimes |\psi\rangle_C\big)\nonumber\\
&&{1\over \sqrt{n_G}}\sum_{\phi\in G}|\phi\rangle_A\otimes |\phi\rangle_B\otimes \big(U(\phi)MU(\phi)^{-1}\big)|\psi\rangle_C~.\nonumber\\
\end{eqnarray}
Furthermore $M^{\rm inv}_{AC}$ and $M^{\rm inv}_{BC}$ act identically on any state of the form
\begin{equation}
|\Psi\rangle_{ABC}= {1\over \sqrt{n_G}}\sum_{\phi\in G}|\phi\rangle_A\otimes |\phi\rangle_B\otimes |\psi_\phi\rangle_C~,
\end{equation}
where $|\psi_\phi\rangle_C$ might depend on $\phi$, a form that is maintained as successive invariant operations are applied to $AC$ and to $BC$.  Therefore, the outcome of the protocol would be the same if each invariant operation $M^{\rm inv}_{BC}$  applied to $BC$ were replaced by the corresponding invariant operation $M^{\rm inv}_{AC}$ applied to $AC$. We conclude that the simulation in which the distributed reference system $AB$ is prepared in the initial state $|0\rangle_{AB}$ is equivalent to a simulation that uses an itinerant reference system $A$. Since this latter simulation has all of the properties listed in Sec.~\ref{subsec:properties}, we find that a bipartite $I$-world protocol using the distributed reference system can faithfully simulate an arbitrary $U$-world protocol.

Note that the distributed state can serve the same purpose if there is a fixed offset of Bob's condensate relative to Alice's, as long as the offset is known. That is, if Alice and Bob share the state
\begin{eqnarray}
&&|0,\tilde\phi\rangle_{AB}= {1\over \sqrt{n_G}}\sum_{\phi\in G}|\phi\rangle_A\otimes |\phi\tilde\phi\rangle_B~\nonumber\\
&&= {1\over\sqrt{n_G}}\sum_{q,a,b}D^q_{ab}(\tilde\phi)\big(\sum_i|\bar q,i,a\rangle_A\otimes |q,i,b\rangle_B\big)~,\nonumber\\
\end{eqnarray}
then the invariant operations $M^{\rm inv}_{BC}$ and $\big(U(\tilde \phi)MU(\tilde \phi)^{-1}\big)_{AC}^{\rm inv}$ act in the same way. If Bob knows $\tilde\phi$, then, he can participate successfully in the simulation by ``twisting'' his operations appropriately.

Similarly, in a protocol with $k$ parties, the distributed reference state
\begin{eqnarray}
|0\rangle_{k {\rm  ~parties}}= {1\over\sqrt{n_G}}\sum_{\phi\in G}|\phi\rangle_{R_1}\otimes |\phi\rangle_{R_2}\cdots \otimes |\phi\rangle_{R_k}
\end{eqnarray}
provides a common ``phase standard'' for all the participants, allowing them to simulate a $U$-world protocol in the $I$-world --- the $\ell$th party simulates the noninvariant operation $M$ by applying $M^{\rm inv}$ to the target system and her part $R_\ell$ of the reference system. Again, the parties can twist their local operations to compensate for known relative offsets of their condensates, if necessary.

In the state $|0\rangle_{AB}$, there is a quantum correlation between Alice's condensate and Bob's. A common reference standard can be provided instead by a classically correlated state such as
\begin{equation}
\rho_{AB}={1\over n_G} \sum_{\phi\in G} \big(|\phi\rangle\langle \phi|\big)_A\otimes \big(|\phi\rangle\langle \phi|\big)_B~.
\end{equation}
If Alice and Bob are equipped with the state $\rho_{AB}$, then again $M^{\rm inv}_{AC}$ and $M^{\rm inv}_{BC}$ act in the same way; hence they can use this distributed reference state to simulate a $U$-world protocol in the $I$-world. The state is $G$-invariant, but unlike $|0\rangle_{AB}$ it is not a charge eigenstate; rather it is a mixture of (invariant) states with various charges. For example, in the case $G=U(1)$, $|0\rangle_{AB}$ is the (unnormalizable) state
\begin{eqnarray}
|0\rangle_{AB}=\int_0^{2\pi}|\theta\rangle_A\otimes |\theta\rangle_B=\sum_{q=-\infty}^{\infty}|-q\rangle_A\otimes |q\rangle_B~;
\end{eqnarray}
Alice's charge and Bob's charge are perfectly anticorrelated. 
In contrast, $\rho_{AB}$ is
\begin{eqnarray}
\rho_{AB}\propto \int d\theta~\left(|\theta\rangle\langle \theta|\right)_A\otimes \left(|\theta\rangle\langle \theta|\right)_B\nonumber\\
\propto \sum_{q_A,q_B,q} |q_A,q_B\rangle\langle q_A-q,q_B+q|~.
\label{mixed_correlated_phase}
\end{eqnarray}
Formally, this state appears to be separable, as it is a mixture of the product states $|\theta\rangle\otimes |\theta\rangle$, but this is deceptive, because $|\theta\rangle\otimes |\theta\rangle$ is not $G$-invariant and is therefore incompatible with the superselection rule. On the other hand, in the charge-eigenstate basis, $\rho_{AB}$ can be expressed as a mixture of $G$-invariant pure states, each with a definite total charge; however, these pure states are highly entangled, with an indefinite value of Alice's (and Bob's) local charge. The state $\rho_{AB}$ is not a mixture of invariant product states, and therefore cannot be prepared without quantum communication between Alice and Bob. Classical communication alone is insufficient for Alice and Bob to establish their common phase standard.

Now let's return to the question we postponed earlier: what if $G$ is a Lie group, so that the states $|0\rangle_{AB}$ and $\rho_{AB}$ are unnormalizable? To be specific, consider again the case $G=U(1)$, and suppose that Alice and Bob are instructed to perform this protocol: Alice is presented with a charge-zero state $|0\rangle$. She is instructed to rotate this state to the superposition of charge eigenstates $\left(|0\rangle + |1\rangle\right)/\sqrt{2}$ and to send the resulting state to Bob. Bob is to perform an orthogonal measurement in the basis $\left(|0\rangle \pm |1\rangle\right)/\sqrt{2}$ and so verify that Alice prepared the correct state. To make sense of this procedure, Alice and Bob must share a common reference state that serves to lock together their phase conventions; for example, this state could be a shared pure state $|\psi\rangle_{AB}$ with definite total charge. Alice's coherent operation on system $C$ acts as
\begin{eqnarray}
&&|\psi\rangle_{AB}\otimes |0\rangle_C\to \nonumber\\
&&{1\over \sqrt{2}}\left(|\psi\rangle_{AB}\otimes |0\rangle_C+ \left(U_-\right)_A|\psi\rangle_{AB}\otimes |1\rangle_C\right)~;
\end{eqnarray}
that is, Alice simulates the charge-nonconserving operator $(U_+)_C$ by applying the invariant operator $(U_-)_A\otimes (U_+)_C$ to $AC$.
When Bob receives system $C$, he performs his measurement by first simulating the transformation
\begin{eqnarray}
&&|0\rangle_C\to {1\over\sqrt{2}}\left(|0\rangle_C+|1\rangle_C\right)~,\nonumber\\
&&|1\rangle_C\to {1\over\sqrt{2}}\left(|0\rangle_C-|1\rangle_C\right)~,
\end{eqnarray}
and then measuring the charge of $C$. After Bob's first step, the state of $ABC$ has become
\begin{eqnarray}
&&{1\over 2}\big(I_A\otimes I_B + (U_-)_A\otimes(U_+)_B\big)|\psi\rangle_{AB}\otimes |0\rangle_C\nonumber\\
&&+{1\over 2}\big(I_A\otimes  (U_-)_B - (U_-)_A\otimes I_B\big)|\psi\rangle_{AB}\otimes|1\rangle_C~.
\end{eqnarray}
When Bob measures the charge, the probability that he obtains the outcome 1 and fails to verify Alice's state, is
\begin{equation}
P_1= {1\over 2}\Big(1 - Re~{}_{AB}\langle\psi|(U_-)_A\otimes(U_+)_B|\psi\rangle_{AB}\Big)~.
\end{equation}
If, for example, the shared reference state is 
\begin{equation}
\label{finite_condensate}
|\psi\rangle_{AB}= {1\over \sqrt{N}}\left(\sum_{q=0}^{N-1}|-q\rangle_A\otimes |q\rangle_B\right)~,
\end{equation}
a normalizable approximation to the state $|0\rangle_{AB}$, our expression for $P_1$ becomes
\begin{equation}
P_1={1\over 2N}~.
\end{equation}
Thus, for finite $N$, the state received by Bob does not match perfectly with the state prepared by Alice --- the superposition of charge eigenstates decoheres slightly. But this decoherence becomes negligible in the limit $N\to \infty$, where the ``charge fluctuations'' of the shared condensate are large.

The lesson we learn from this example generalizes to nonabelian compact Lie groups. We can replace the unnormalizable state
\begin{equation}
|0\rangle_{AB}={1\over\sqrt{n_G}}\sum_{q,i,a}|\bar q,i,a\rangle_A\otimes |q,i,a\rangle_B
\end{equation}
by a normalizable state with a truncated sum over the charge $q$. If Alice and Bob use this truncated distributed reference state to simulate a $U$-world protocol, their simulation will not have perfect fidelity. But as long as all operations applied by Alice and Bob change the charge by a bounded amount, the fidelity can be arbitrarily close to one if the reference state is chosen appropriately. If Alice and Bob are permitted to use a truncated {\em itinerant} reference system rather than a distributed one, then perfect fidelity can be achieved, as observed in Sec.~\ref{subsec:properties}.

\section{Invariant operations and commutants}
\label{sec:commutants}

Our observations in Sec.~\ref{subsec:nonabelian} emphasized the similarities between abelian and nonabelian superselection rules, enabling us to formulate a security analysis in Sec.~\ref{sec:simulation} that applies to both abelian and nonabelian symmetry groups. But in several respects the arguments in Sec.~\ref{sec:simulation} are still not adequate. For one thing,  so far we have treated only the special case of superselection sectors labeled by unitary irreducible representations of compact groups. For another, while it is possible to formulate a security analysis of quantum bit commitment within the framework of our argument in Sec.~\ref{sec:simulation}, it is more natural to structure the argument differently, following more closely the standard analysis of quantum bit commitment.

In this section, we will emphasize the essential differences between superselection rules arising from nonabelian symmetry groups and those arising from abelian groups. The discussion will pave the way for our analysis of quantum bit commitment in Sec.~\ref{sec:qbc} and of general two-party protocols in Sec.~\ref{sec:two-party}. 

A crucial difference between abelian and nonabelian charges is that nonabelian charges are nonadditive: the charges of two subsystems $A$ and $B$ do not necessary determine the charge of the composite system $AB$. This feature can be restated as a property of the algebra of observables of the bipartite system.  Let $\cal A$ denote the algebra of local operators (an associative algebra, closed under Hermitian conjugation, that commutes with all locally conserved charges) acting on subsystem $A$, and let $\cal B$ denote the algebra of local operators acting on $B$. The commutant of $\cal A$, denoted $\cal A'$, is the algebra of operators acting on the composite system $AB$ that commute with everything in $\cal A$, and similarly for $\cal B'$. Now, if all superselection rules are abelian, then $\cal A'=\cal B$ and $\cal B'=\cal A$. But if the superselection rules are nonabelian, the theory has sectors with nontrivial total charge in which this relation does not hold.  This unusual structure of the local observables has potential implications for the security of quantum protocols.

To be more explicit, suppose that the superselection rules arise from a nonabelian symmetry group $G$, and the operations that Alice (or Bob) can perform must commute with $G$. A state $|\psi\rangle$ in Alice's (or Bob's) Hilbert space can be decomposed into irreducible representations of $G$, as 
\begin{equation}
|\psi\rangle = \sum_{q,i,a} \psi^q_{i,a} |q,i,a\rangle ~;
\end{equation}
here $q$ labels the irreducible representation (or ``charge''), $i$ is the ``color'' index acted upon by the representation of $G$, and $a$ is the ``flavor'' index that distinguishes among the various copies of the irreducible representation $q$ appearing in the decomposition. Note that since we are no longer assuming that Alice's system transforms as the regular representation of $G$, there need be no connection between the number of flavors and the number of colors associated with $q$. The action of a color gauge rotation representing $g\in G$ on $|\psi\rangle$ is
\begin{equation}
U(g)|\psi\rangle =  \sum_{q,i,j,a} \psi^q_{i,a} |q,j,a\rangle D^{q}_{ji}(g)~.
\end{equation}
An operator $M$ allowed by the superselection rule, which must commute with each $D^{q}(g)$, preserves the charge $q$ and acts only on the flavor index according to
\begin{equation}
M|\psi\rangle =  \sum_{q,i,a,b} \psi^q_{i,a} |q,i,b\rangle M^{q}_{ba}~.
\end{equation}
Since allowed operations act nontrivially only on the flavor index, it is convenient to use a notation that suppresses the color index $i$. We denote by ${\cal H}_q$ the {\em invariant} Hilbert space in the charge-$q$ sector, spanned by states $|q,a\rangle$ that are labeled only by the flavor $a$ within the sector. The corresponding operator algebra respecting the superselection rule is ${\cal L}({\cal H}_q)$, spanned by linear operators acting on this invariant space. Thus Alice's invariant Hilbert space is
\begin{equation}
{\cal H}_A= \bigoplus_q~ {\cal H}_{A,q}
\end{equation}
and  Alice's local operator algebra is
\begin{equation}
{\cal A}= \bigoplus_q ~ {\cal L}({\cal H}_{A,q})~;
\end{equation}
Similarly, Bob's operator algebra is 
\begin{equation}
{\cal B}= \bigoplus_q ~ {\cal L}({\cal H}_{B,q})~.
\end{equation}

Now consider the composite system $AB$. Its invariant Hilbert space too can be expressed as a direct sum over charge sectors \begin{equation}
{\cal H}= \bigoplus_q {\cal H}_q~,
\end{equation}
while the  full operator algebra is $\oplus_q{\cal L}({\cal H}_q)$. But we should consider how ${\cal H}_q$ is related to the invariant Hilbert spaces of the subsystems.  The charge-$q$ Hilbert space of the joint system can be expressed as
\begin{equation}
{\cal H}_q= \bigoplus_{q_A,q_B} {\cal H}_{A,q_A}\otimes {\cal H}_{B,q_B}\otimes V_q^{q_A,q_B}~,
\label{hilbert_expand}
\end{equation}
where $V_q^{q_A,q_B}$ denotes the space of invariant linear maps from the irreducible representation $q$ to the tensor product of irreducible representations $q_A\otimes q_B$. This space can be nontrivial (of dimension greater than one) if the tensor product contains the representation $q$ more than once.

When expressed in terms of a particular color basis for the irreducible representations $q$, $q_A$ and $q_B$, the components of 
$V_q^{q_A,q_B}$ are the Clebsch-Gordon coefficients ($3j$ symbols), of the group $G$. Let $\{|q_A, i\rangle\}$ denote an orthonormal basis for the representation $q_A$, $\{|q_B, j\rangle\}$ a basis for  $q_B$, and $\{|q(\alpha), k\rangle\}$ a basis for  $q(\alpha)$, where the index $\alpha$ labels the various copies of the representation $q$ that may be contained in $q_A\otimes q_B$. Then the components of $V_q^{q_A,q_B}$ are
\begin{eqnarray}
\big[V_q^{q_A,q_B}(\alpha)\big]_k^{ij}= \big(\langle q_A,i|\otimes \langle q_B,j|\big)|q(\alpha),k\rangle~. 
\end{eqnarray}
These components comprise a $G$-invariant tensor with the property
\begin{eqnarray}
&&\big[V_q^{q_A,q_B}(\alpha)\big]_k^{ij}\nonumber\\
&&=\sum_{i',j',k'}D^{q_A}_{ii'}(g)D^{q_B}_{jj'}(g)\big[V_q^{q_A,q_B}(\alpha)\big]_{k'}^{i'j'}D^{q}_{k'k}(g)~.
\end{eqnarray}
Invariant operations act not on the color indices of $\big[V_q^{q_A,q_B}(\alpha)\big]_k^{ij}$, but rather on the index $\alpha$ that distinguishes the flavors of $q$ contained in $q_A\otimes q_B$. Furthermore, the invariant operations can also alter the charges $q_A$ and $q_B$ appearing in eq.~(\ref{hilbert_expand}), while preserving the total charge $q$.

The notation of eq.~(\ref{hilbert_expand}) and its implications may be clarified by discussing  specific examples. The trivial representation $(q=1)$ is contained only in the tensor product of $q_A$ with its conjugate representation $\bar q_A$, and it occurs only once in this product; Therefore, in the case where the total charge is $q=1$, eq.~(\ref{hilbert_expand}) reduces to
\begin{equation}
{\cal H}_1= \bigoplus_{q} {\cal H}_{A, q}\otimes {\cal H}_{B, \bar q} ~;
\end{equation}
in this case, the factor $V_q^{q_A,q_B}$ is superfluous. Now, the joint operator algebra contains operations that cannot be executed by Alice and Bob locally --- these operations change Alice's charge and Bob's while preserving the total charge (of course, this can happen even if $G$ is abelian). But any operation that commutes with Alice's algebra ${\cal A}$ must preserve Alice's charge $q$, and act trivially in each of Alice's charge sectors; such operations preserve Bob's charge $\bar q$ as well, and thus are in Bob's algebra ${\cal B}$. Therefore ${\cal A}$ and ${\cal B}$ are commutants of one another.

However, if the total charge is nontrivial, then ${\cal B}$ need not be the commutant of ${\cal A}$. To illustrate this phenomenon, consider the case $G=SU(2)$, where the irreducible representation is labeled by the spin $j$. For $SU(2)$, $V_j^{j_A,j_B}$ is always one (or zero) dimensional, and eq.~(\ref{hilbert_expand}) reduces to 
\begin{equation}
{\cal H}_j= \bigoplus_{j_A,j_B} {\cal H}_{A,j_A}\otimes {\cal H}_{B,j_B}~,
\end{equation}
where it is implicit that each product of representations appearing on the right-hand side transforms as spin $j$. To be concrete, suppose that Alice's  system has spin $1/2$,  Bob's contains both a spin-$0$ and a spin-$1$ component, and the total spin is $1/2$; then
\begin{equation}
{\cal H}_{1/2}= {\cal H}_{A,1/2}\otimes \left({\cal H}_{B,0}\oplus {\cal H}_{B,1}\right)~.
\label{su2_example}
\end{equation}
Note that in this case, contrary to the case in which the total charge is trivial, a single value of $j_A$ can be combined with either of two different values of $j_B$ to obtain the same total charge $j$. Therefore, there are invariant operations acting on the joint system that preserve Alice's charge and the total charge, but change Bob's charge. These operations are in the commutant of ${\cal A}$ but not in ${\cal B}$; hence ${\cal A}' \ne {\cal B}$. 

We arrive at another way of looking at this property of ${\cal H}_{1/2}$ if we imagine that there is a third party Charlie who holds a compensating charge, so that the total charge is trivial. Now
\begin{equation}
{\cal H}_{0}= {\cal H}_{A,1/2}\otimes \left({\cal H}_{B,0}\otimes {\cal H}_{C,1/2} \oplus {\cal H}_{B,1}\otimes {\cal H}_{C,1/2}\right)~;
\label{ABC_system}
\end{equation}
an operation in ${\cal A}'$ can be performed by Bob and Charlie acting together, but not by Bob alone.

In order that ${\cal A}'\ne {\cal B}$, it is not necessary for one of the parties to possess a state with indefinite charge. For example, in the case $G=SU(3)$, the tensor product of the irreducible octet representation $8$ with itself contains two copies of $8$, one symmetric and one antisymmetric under interchange of the factors:
\begin{equation}
8_A\otimes 8_B \supseteq 8_{\rm sym} \oplus 8_{\rm anti}~.
\end{equation}
Thus, in the decomposition
\begin{equation}
{\cal H}_8= {\cal H}_{A,8}\otimes {\cal H}_{B,8}\otimes V_8^{8,8}~,
\label{su3_example}
\end{equation}
the joint invariant Hilbert space is two-dimensional, while Alice and Bob both have one-dimensional Hilbert spaces and trivial invariant operator algebras. Then  ${\cal A}'$ is the full operator algebra, clearly different from ${\cal B}$, and similarly ${\cal B}'$ is different from ${\cal A}$. Again, an alternative description of the invariant space is to note that Charlie could hold a compensating 8 charge, in which case the total charge is trivial and
\begin{equation}
{\cal H}_1= \left({\cal H}_{A,8}\otimes {\cal H}_{B,8}\otimes {\cal H}_{C,8}\right)\otimes V_1^{8,8,8}
\end{equation}
is two-dimensional.

For the purpose of describing $G$-invariant operations, it is always legitimate to introduce a compensating charge without incurring any loss of generality. To see this, first note that if ${\cal E}$ is a $G$-invariant quantum operation, then
\begin{eqnarray}
{\cal E}\big[U(g)\rho U(g)^{-1}\big]= U(g) {\cal E}(\rho) U(g)^{-1}
\end{eqnarray}
for any $g\in G$ and any state $\rho$. In particular, then, 
\begin{eqnarray}
\label{EandGcommute}
{\cal E}\big[{\cal G}(\rho)\big]= {\cal G}\big[{\cal E}(\rho)\big]~,
\end{eqnarray}
where ${\cal G}$ is the map
\begin{eqnarray}
{\cal G}(\rho)={1\over n_G}\sum_{g\in G} U(g)\rho U(g)^{-1}~,
\end{eqnarray}
which induces decoherence of a superposition of distinct irreducible representations of $G$:
\begin{eqnarray}
&&{\cal G}\big(|q,i,a\rangle\langle q',j,b|\big)\nonumber\\
&&=\delta^{qq'}\delta_{ij}\left({1\over n_q}\sum_l|q,l,a\rangle\langle q,l,b|\right).\end{eqnarray}
Eq.~(\ref{EandGcommute}) means \cite{bartlett} that the state
\begin{eqnarray}
|\psi\rangle=\sum_{i,a} \psi^q_{i,a}|q,i,a\rangle
\end{eqnarray}
cannot be distinguished by any $G$-invariant operation from the state
\begin{eqnarray}
\label{state_Ginvariant}
{\cal G}\big(|\psi\rangle\langle \psi|\big)
=\sum_{q,a,b,i}\psi^q_{i,a}\psi^{q~*}_{i,b}\left({1\over n_q}\sum_j |q,j,a\rangle\langle q,j,b|\right).
\end{eqnarray}

Now, consider a system $A$ whose charge is screened by a system $C$, so that the state of the joint system has trivial total charge:
\begin{eqnarray}
|\psi\rangle_{AC}=\sum_{q,a,i}\psi^q_a~|q,i,a\rangle_A\otimes |\bar q,i\rangle_C~.
\end{eqnarray}
Tracing over system $C$ produces the state
\begin{eqnarray}
\label{traceout}
{\rm tr}_C\big(|\psi\rangle\langle \psi|\big)_{AC}= \sum_{q,a,b}\psi^q_{a}\psi^{q~*}_{b}\left({1\over n_q}\sum_j |q,j,a\rangle\langle q,j,b|\right)~.
\end{eqnarray}
But the state eq.~(\ref{state_Ginvariant}) is just a convex combination of states of the form eq.~(\ref{traceout}). Therefore, if only $G$-invariant operations are to be considered, it is always harmless to replace system $A$ by half of a bipartite state that carries trivial total charge.

Up until now, we have explicitly discussed only the case of superselection sectors arising from a compact symmetry group, but much of the formalism we have outlined in this section can be extended to a more general setting. Whatever the origin of the superselection rule, the allowed operations act on a suitable invariant space. Sectors can still be classified by conserved charges, but in the general case, the space $V_q^{q_A,q_B}$ is defined more abstractly, rather than in terms of group representations. One important property that continues to hold in the general setting (which will play a central role in our analysis of quantum bit commitment in Sec.~\ref{sec:qbc} and of general two-party games in Sec.~\ref{sec:two-party}) is that for each value $q$ of the charge, there is a unique conjugate charge $\bar q$ such that the fusion of the charges contains the trivial charge sector.

\section{Data hiding}
\label{sec:data}
Verstraete and Cirac \cite{verstraete} described a data-hiding protocol whose security is founded on the charge superselection rule for $G=U(1)$. Suppose that a trusted third party Charlie prepares one of the two orthogonal states
\begin{equation}
|\pm\rangle = {1\over \sqrt{2}}\left(|01\rangle \pm |10\rangle\right)~,
\label{verstrate_states}
\end{equation}
where $|0\rangle$ and $|1\rangle$ denote states of charge 0 and 1 respectively, and distributes half to Alice and half to Bob. If Alice and Bob could each measure the Pauli operator $X$ that interchanges $|0\rangle$ and $|1\rangle$, they could distinguish the states $|+\rangle$ and $|-\rangle$ by performing these measurements and comparing their outcomes. However, $X$ does not commute with the electric charge $Q$; if Alice and Bob are permitted only to perform local charge-conserving operations and to communicate classically, then they will be powerless to distinguish the two possible states.

On the other hand, if Alice and Bob share access to a common phase reference state, their activities will be unrestricted and nothing will prevent them from performing the $X$ measurements that unlock the classical bit stored in the state prepared by Charlie (aside from the small loss of fidelity that arises if the reference state has large but finite charge fluctuations, as in eq.~(\ref{finite_condensate})). In Bloch sphere language, Alice and Bob have no {\it a priori} means of orienting their measurement axes in the $x$-$y$ plane, but a shared phase standard enables them to lock their axes together and compare their measurements. Since the state prepared by Charlie is invariant under rotations about the $z$ axis, the overall orientation in the $x$-$y$ plane is irrelevant; only the relative orientation needs to be fixed to identify Charlie's state.

To be more explicit, while $X$ does not commute with the charge, 
\begin{equation}
X^{\rm inv}_{AA'} =\left(U_-\right)_A \otimes\sigma^+_{A'} + \left(U_+\right)_A \otimes\sigma^-_{A'}
\end{equation}
commutes with $Q$, as does $X^{\rm inv}_{BB'}$. If Alice and Bob share a distributed reference state $|\psi\rangle_{AB}$ that is an eigenstate of $\left(U_-\right)_A\otimes \left(U_+\right)_B$ with eigenvalue $1$, then
\begin{equation}
|\psi\rangle_{AB}\otimes |\pm\rangle_{A'B'}
\end{equation}
is an eigenstate of 
\begin{equation}
X^{\rm inv}_{AA'}\otimes X^{\rm inv}_{BB'}
\end{equation}
with eigenvalue $\pm 1$.
Therefore, Alice and Bob can unlock the hidden bit by each measuring $X^{\rm inv}$ and comparing their results.
The same holds, of course, if the shared reference state $\rho_{AB}$ is a mixture of eigenstates of $\left(U_-\right)_A\otimes \left(U_+\right)_B$, each with eigenvalue 1, as in eq.~(\ref{mixed_correlated_phase}). As Verstraete and Cirac observed \cite{verstraete}, quantum communication is needed to establish this shared phase standard.

In the absence of a shared phase standard, neither Alice nor Bob can detect the bit encoded in the state $|\pm\rangle$ of eq.~(\ref{verstrate_states}); however, either Alice or Bob can manipulate the bit. Each can measure the charge $q$, and either can apply a phase to the state conditioned on the charge, flipping $|+\rangle\leftrightarrow |-\rangle$. But the property that ${\cal B}'\ne{\cal A}$ indicates that the situation can be more subtle in the nonabelian case (with nontrivial total charge). Suppose, for example, that $G=SU(2)$ with total charge $j=1/2$ as in eq.~(\ref{su2_example}). Two states with the same value of the total charge and of Alice's charge, but different values of Bob's charge, are $|j=1/2, j_A=1/2, j_B=0\rangle$ and $|j=1/2, j_A=1/2, j_B=1\rangle$. Charlie might prepare either of the linear combinations
\begin{eqnarray}
|\pm\rangle = {1\over\sqrt{2}} \Big(\big|j={1\over 2}, ~j_A={1\over 2}, ~j_B=0\big\rangle\nonumber\\
\pm ~\big|j={1\over 2}, ~j_A={1\over 2}, ~j_B=1\big\rangle\Big)~,
\label{two_su2_states}
\end{eqnarray}
and then distribute the $AB$ system to Alice and Bob.  Again, neither Alice nor Bob can detect the hidden bit, but now there is a notable asymmetry between Alice's power and Bob's. Since Bob has a superposition of two different charge states, he can tamper with the hidden bit by applying a phase controlled by the charge. Alice, on the other hand, has a trivial invariant operator algebra, and has no control over the shared state.

We may take this observation a step further. Suppose, for example, that $G=SU(3)$ with total charge $q=8$ as in eq.~(\ref{su3_example}). Charlie might prepare either of the linear combinations
\begin{eqnarray}
|\pm\rangle = {1\over\sqrt{2}} \Big(\big|q=8_{\rm sym}, ~q_A=8, ~q_B=8\big\rangle\nonumber\\
\pm ~\big|q=8_{\rm anti}, ~q_A={8}, ~q_B=8\big\rangle\Big)~,
\end{eqnarray}
and then distribute the $AB$ system to Alice and Bob.  Again, neither Alice nor Bob can detect the hidden bit, but furthermore, neither one can tamper with the bit's value.

However, in the nonabelian case as in the abelian case, the hidden bit can be opened via local operations and classical communication between Alice and Bob if they are provided with correlated reference systems that effectively remove the restrictions imposed by the superselection rule.

\section{Quantum bit commitment and superselection rules}
\label{sec:qbc}
During the commitment stage of  quantum bit commitment, Alice encodes a classical bit by preparing one of two distinguishable quantum states with density operators $\rho_0$ or $\rho_1$, and then she sends half of the state to Bob. In the unveiling stage, Alice sends the other half of the state to Bob, so that he can verify whether the state is $\rho_0$ or $\rho_1$. The protocol is binding if, after commitment, Alice is unable to change the value of the bit. The protocol is concealing if, after commitment and before unveiling, Bob is unable to discern the value of the bit. The protocol is secure if it is both binding and concealing.

In the absence of superselection rules, unconditionally secure quantum bit commitment is impossible \cite{mayers_qbc,lo_chau}. If we imagine that the states $\rho_0$ and $\rho_1$ are pure states shared by Alice and Bob, then if the protocol is concealing, Bob's density operator (obtained by tracing over Alice's system) must be the same in both cases: $\rho_{0,B}=\rho_{1,B}$. But then by the HJW Theorem \cite{hjw} Alice can apply a unitary transformation to her half of the state that transforms $\rho_0$ to $\rho_1$, so that the protocol is not binding. 

\subsection{Bit commitment with mixed states}
\label{qbc:mixed}
We reached this conclusion under the assumption that $\rho_0$ and $\rho_1$ are pure states, but we can extend the argument to the case were the states are mixed by appealing to the concept of a {\em purification} of a mixed state. We will describe this extension in detail, as we will follow very similar reasoning in our discussion in Sec.~\ref{subsec:nontrivial_charge} of bit commitment with nontrivial total charge.

Suppose that at the start of the bit commitment protocol, Alice and Bob share a product state $\rho_A\otimes \rho_B$, where the states $\rho_A$ and $\rho_B$ are mixed. An equivalent way to describe Alice's initial state is to introduce the ancilla system $C$ and a pure state $|\psi\rangle_{AC}$ (a purification of $\rho_A$), such that the density operator $\rho_A$ is obtained from $|\psi\rangle_{AC}$ by tracing over system $C$:
\begin{equation}
\rho_A={\rm tr}_C\big( |\psi\rangle\langle\psi|\big)_{AC}~.
\end{equation}
Similarly, to describe $\rho_B$ we can introduce the ancilla $D$ and a state $|\varphi\rangle_{BD}$ that purifies $\rho_B$. Without loss of generality, we may assume that in each step of the protocol, Alice or Bob applies a unitary transformation, so that the state of the full system $ABCD$ remains pure. (A general quantum operation performed by Alice, say, can be realized as a unitary transformation applied jointly to Alice's system and to an appropriate ancilla; therefore, the operation is unitary provided that we include this ancilla as part of the system.) In particular, after the bit is committed, the state of the full system is one of the two pure states $|\psi_0\rangle_{ABCD}$ or $|\psi_1\rangle_{ABCD}$.

If both parties are honest, the ancillas $C$ and $D$ are off limits --- Alice can manipulate only $A$ and Bob can manipulate only $B$ --- and in that case the mixed state protocol and its purification are completely equivalent. Furthermore, if one party cheats, whether the other party starts out with a mixed state or its purification has no impact on the effectiveness of the cheating strategy, because the honest party never touches the purifying ancilla anyway.

Now let us see that in any quantum bit commitment protocol, one of the players can cheat successfully. First suppose that Bob cheats. Though the honest protocol calls for Bob to start our with the mixed state $\rho_B$, a cheating Bob can throw this state away, and replace it with the purification $|\varphi\rangle_{BD}$, where $D$ is now {\em an ancilla system that Bob controls}. Therefore, if the protocol is perfectly concealing (even when Bob cheats), then 
\begin{eqnarray}
&&\rho_{0,BD}\equiv {\rm tr}_{AC} \big(|\psi_0\rangle\langle \psi_0|\big)_{ABCD}\nonumber\\
&&=\rho_{1,BD}\equiv {\rm tr}_{AC} \big(|\psi_1\rangle\langle \psi_1|\big)_{ABCD}~;
\end{eqnarray}
Bob is unable to collect any information about the committed bit through any joint measurement on $BD$.

Similarly, a cheating Alice could throw away her initial state and replace it by its purification; then Alice could control both $A$ and the ancilla $C$. Applying the HJW theorem as before, we conclude that if $\rho_{0,BD}=\rho_{1,BD}$, then Alice can apply a unitary transformation to $AC$ that transforms $|\psi_0\rangle_{ABCD}$ to $|\psi_1\rangle_{ABCD}$. We conclude that if the protocol is concealing, then it is not binding. Unconditionally secure quantum bit commitment is impossible, even with mixed states. That quantum bit commitment is impossible even when mixed strategies are used was proved in \cite{mayers_qbc} using a slightly different approach. 

\subsection{Trivial total charge}
\label{subsec:qbc-trivial}
The argument in Sec.~\ref{qbc:mixed} shows that for an analysis of the security of quantum bit commitment, we may assume that Alice and Bob share a pure state. But how is the security affected if superselection rules constrain Alice's and Bob's operations? We will first consider the special case in which the total charge that Alice and Bob share is trivial. 
After commitment, then, Alice and Bob share one of the two pure states $|\psi_0\rangle$ or $|\psi_1\rangle$, each with trivial total charge. Choosing the Schmidt basis in each charge sector, the state $|\psi_0\rangle$ can be expanded as
\begin{equation}
|\psi_0\rangle_{AB}= \sum_q \sqrt{p_q}\sum_b \sqrt{\lambda_{q,b}} ~|\bar q,b\rangle_A\otimes |q,b\rangle_B~.
\end{equation}
where Bob's density operator is
\begin{equation}
\rho_{0,B}={\rm tr}_A \left(|\psi_0\rangle\langle \psi_0|\right) = \sum_q p_q ~\rho_{0,B,q}
\end{equation}
and
\begin{equation}
\rho_{0,B,q}= \sum_b \lambda_{q,b} ~|q,b\rangle\langle q,b|~.
\end{equation}
Bob can measure the probability $p_q$ that his charge is $q$; therefore if the protocol is concealing then the distribution $\{p_q\}$ must be the same for $|\psi_1\rangle$ as for $|\psi_0\rangle$. Furthermore, Bob's density operator in the charge-$q$ sector must not depend on whether the state is $|\psi_0\rangle$ or $|\psi_1\rangle$; therefore $|\psi_1\rangle$ can be expanded as 
\begin{equation}
|\psi_1\rangle_{AB}= \sum_q \sqrt{p_q}\sum_b \sqrt{\lambda_{q,b}} ~|\bar q,\tilde b\rangle_A\otimes |q,b\rangle_B~,
\end{equation}
where $\{|\bar q,\tilde b\rangle_A\}$ is another basis for Alice's charge-$\bar q$ sector. But now Alice can apply a unitary transformation conditioned on the charge that rotates one basis to the other:
\begin{equation}
U_{\bar q}: |\bar q,b\rangle \to |\bar q,\tilde b\rangle~,
\end{equation}
which transforms $|\psi_0\rangle$ to $|\psi_1\rangle$. Therefore, the protocol is not binding.

Obviously, the same argument applies, in the abelian case, even if the total charge is nontrivial \cite{mayers}. The key property of the states that is used in the argument is that Alice's charge is perfectly correlated with Bob's, so that ${\cal B}'={\cal A}$.

\subsection{Nontrivial total charge}
\label{subsec:nontrivial_charge}

The property that ${\cal B}'\ne{\cal A}$ in the nonabelian case (with nontrivial total charge) encourages one to hope that a  bit commitment protocol can be formulated whose security is founded on a nonabelian superselection rule. Indeed, consider again the case $G=SU(2)$ with total charge $j=1/2$ as in eq.~(\ref{su2_example}). When Alice has control of the full $AB$ system, she can prepare either of the states $|\pm\rangle_{AB}$ shown in eq.~(\ref{two_su2_states}),
and then she can send the $B$ system to Bob. Now Bob is unable to distinguish the two states, because he cannot measure the relative phase in a superposition of two states of different charge.  Furthermore there is no invariant operation Alice can apply that changes $|+\rangle$ to $|-\rangle$ or vice versa. It seems, then, that the protocol is both concealing and binding!
At any rate, quantum bit commitment in a world with nonabelian superselection rules seems fundamentally different than quantum bit commitment in a world in which all superselection rules are abelian.

But, as always in a discussion of information-theoretic security, we must be sure to consider the most general possible cheating strategies. And in fact, we can argue that for the security analysis, there is no loss of generality if we assume that the charge shared by the parties is trivial, the case we have already dealt with in Sec.~\ref{subsec:qbc-trivial}.  This reduction to the case of trivial total charge follows closely our discussion in Sec.~\ref{qbc:mixed}, where we showed that it suffices to assume that the parties share a pure state.

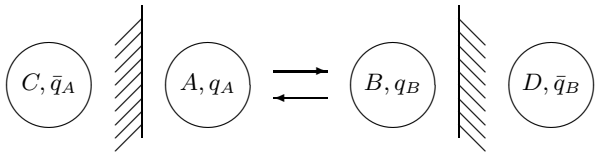
\begin{figure}
\centering
\begin{picture}(240,60)

\put(20,20){\circle{30}}
\put(20,20){\makebox(0,0){$C, \bar q_A$}}

\put(80,20){\circle{30}}
\put(80,20){\makebox(0,0){$A, q_A$}}

\put(55,0){\line(0,1){50}}

\put(55,45){\line(-1,-1){10}}
\put(55,40){\line(-1,-1){10}}
\put(55,35){\line(-1,-1){10}}
\put(55,30){\line(-1,-1){10}}
\put(55,25){\line(-1,-1){10}}
\put(55,20){\line(-1,-1){10}}
\put(55,15){\line(-1,-1){10}}
\put(55,10){\line(-1,-1){10}}
\put(55,5){\line(-1,-1){10}}

\put(105,25){\vector(1,0){20}}
\put(125,15){\vector(-1,0){20}}

\put(150,20){\circle{30}}
\put(150,20){\makebox(0,0){$B,  q_B$}}

\put(210,20){\circle{30}}
\put(210,20){\makebox(0,0){$D, \bar q_B$}}

\put(175,0){\line(0,1){50}}

\put(175,45){\line(1,-1){10}}
\put(175,40){\line(1,-1){10}}
\put(175,35){\line(1,-1){10}}
\put(175,30){\line(1,-1){10}}
\put(175,25){\line(1,-1){10}}
\put(175,20){\line(1,-1){10}}
\put(175,15){\line(1,-1){10}}
\put(175,10){\line(1,-1){10}}
\put(175,5){\line(1,-1){10}}

\end{picture}
\caption{``Purification'' of a two-party game with nontrivial total charge. At the beginning of the game, the charge of $C$ (hidden behind a brick wall) compensates for Alice's charge $q_A$, and the charge of $D$ (also hidden) compensates for Bob's charge $q_B$. Honest players never touch the compensating charges, but a cheating Alice might manipulate $C$ and a cheating Bob might manipulate $D$.}
\label{fig:compensating_charge}
\end{figure}

Consider a general two-party quantum bit commitment protocol in which the initial state shared by Alice and Bob is a tensor product $\rho_A\otimes\rho_B$ of invariant states. The state $\rho_A$ can be purified if we introduce an ancilla $C$; furthermore, the pure state of $AC$ can be chosen to have trivial total charge. Similar, we can purify $\rho_B$ using the ancilla $D$, in such a way that the pure state of $BD$ has trivial total charge. (See Fig.~\ref{fig:compensating_charge}.) Each operation performed by Alice or Bob can be taken to be a charge-conserving unitary transformation; therefore, at each stage of the protocol, the state of the full system $ABCD$ is a pure state with trivial total charge.

In the honest protocol, the ancillas $C$ and $D$ are inaccessible. But if Bob cheats, he can throw away the initial invariant state $\rho_A$ called for in the protocol, and replace it by a trivially charged pure state of $BD$, where $D$ is now an ancilla that Bob controls. Therefore, if the bit commitment protocol is concealing, then $\rho_{0,BD}=\rho_{1,BD}$ --- Bob can't learn anything about the committed bit from any invariant joint measurement on $BD$. Since the state of the full system $ABCD$ is a pure state with trivial charge, the argument of Sec.~\ref{subsec:qbc-trivial} suffices to show that Alice can transform $|\psi_0\rangle$ to $|\psi_1\rangle$ with an invariant local operation applied to $AC$. Hence, the protocol is not binding. We have proved, then, that, even when the protocol calls for a nontrivial total charge, if Bob is unable to cheat then Alice can cheat --- unconditionally secure quantum bit commitment is impossible. We have:

\medskip
\noindent{\bf Theorem 2} {\em Consider a quantum bit commitment protocol in the $I$-world, where at the beginning of the protocol Alice and Bob share a product of invariant states. Then if the protocol is concealing, it is not binding.}
\medskip

\noindent Our proof, which reduces the case of nontrivial total charge to the case of trivial total charge, is really just a minor variant of the argument in Sec.~\ref{qbc:mixed} that reduces the case of a protocol where Alice and Bob share a mixed state to the case where they share a pure state.

In the case of our bit commitment protocol in which the total charge of $AB$ is $j=1/2$, if Alice is unable to access the compensating charge in $C$, then she can't cheat successfully. But if Alice controls the whole $AC$ system, then Alice's charge $j_{AC}=0,1$ is perfectly correlated with Bob's, and she can rotate the relative phase of the $j_{AC}=0$ and $j_{AC}=1$ components of her state, transforming $|+\rangle$ to $|-\rangle$.

This reduction of a protocol with nontrivial total charge to a protocol with trivial total charge can be generalized. In the $I$-world, consider an $n$-party protocol in which up to $k<n$ of the parties might cheat, where the initial state is the product of invariant states $\otimes_{i=1}^n\rho_i$, and where all operations performed by the parties are required to conserve the local charge. Then we may imagine that each party is issued a compensating charge at the beginning of the protocol, so that each party actually starts out with trivial charge. The honest parties will never touch their compensating charges, but a cheating party cannot be prevented from performing arbitrary joint operations on her system and her compensating charge. This strategy is realizable because the cheater might throw away the invariant state she holds at the beginning of the protocol, and replace it by a charge-zero state that she controls fully. Furthermore, if an attack by the cheaters is successful in the protocol where the honest players start out with trivial charge, then it will also be successful if the honest players start out with a product of charged invariant states;  since honest players never make use of the compensating charges, their presence can have no impact on the effectiveness of the attack. Therefore, we have:

\medskip\medskip\medskip
\noindent{\bf Theorem 3} {\em Let $P$ be an $n$-party quantum protocol in the $I$-world that securely realizes a task $\Pi$, where the initial state in $P$ is a product of $n$ invariant states. Then there is an $I$-world protocol $P'$ that also securely realizes $\Pi$, where the initial state in $P'$ is a product of $n$ pure states, each with trivial charge.}

\medskip

\noindent In other words, in a security analysis, we may assume without any loss of generality that each party holds a pure state with trivial charge at the start of the protocol. 

Note that for the proofs of Theorems 2 and 3, our observations from Sec.~II and III on the use of reference systems are not needed. Rather, to prove Theorems 2 and 3, we use only two properties of the $I$-world superselection sectors: first, that for each charge sector ${\cal H}_q$ there is a unique conjugate charge sector ${\cal H}_{\bar q}$ such that the trivial sector ${\cal H}_1$ is contained in ${\cal H}_q\otimes {\cal H}_{\bar q}$, and second, that any invariant state has a purification with trivial total charge. These properties hold not just for the case of superselection rules arising from a symmetry group $G$, but also for the more general superselection rules considered in Sec.~\ref{sec:two-party}. Therefore, Theorems 2 and 3 apply in this more general setting.

\section{Two-party protocols in general}
\label{sec:two-party}

\subsection{Overview}
\label{subsec:overview}

We will now analyze the impact of superselection rules on the security of general two-party protocols. We will show that for any protocol $P$ in the invariant world ($I$-world) subject to the superselection rule, there is a corresponding protocol $\simu{P}$ in the unrestricted world ($U$-world), where $\simu{P}$ {\em simulates} $P$ in the following sense: First, when performed honestly, $\simu{P}$ and $P$ accomplish the same task. And second, for any cheating strategy that can be adopted by a dishonest party in $\simu{P}$, there is a corresponding cheating strategy in $P$ that is just as effective. In particular then, if $\simu{P}$ is insecure, then so is $P$. We conclude, therefore, that superselection rules cannot enhance the (information-theoretic) security of two-party protocols. The methods we will use to establish this result are quite different than those used in Sec.~\ref{sec:simulation} to treat the case of superselection rules arising from a symmetry group.

Before going into the details, we will briefly describe the main ideas used in our argument. First of all, we will restrict out attention to a protocol in which the total charge shared by the two parties is trivial (belongs to the trivial superselection sector). We know from Theorem 3 in Sec.~\ref{subsec:nontrivial_charge} that it suffices to treat this special case in an analysis of security. A protocol with trivial total charge has this useful property: if Alice knows that she holds charge $q$ after sending a message to Bob, then Alice also knows that Bob will hold the conjugate charge $\bar q$ upon receiving the message. Similarly, Bob knows what Alice's charge will be after she receives a message sent by Bob. Our analysis of security relies on the property that Bob has a definite charge if Alice does, and therefore it applies only to two-party protocols. 

In the $I$-world, charge is conserved, so that the total charge shared by Alice and Bob is trivial at each stage of the protocol; furthermore, local operations performed by Alice or Bob must preserve the conserved charge. In the $U$-world, charge need not be conserved, but the protocol $\simu{P}$ that simulates the $I$-world protocol $P$ can be chosen to respect conservation of a fictitious ``charge'' that behaves like the actual conserved charge of the $I$-world. However, a dishonest party who is not bound to follow the protocol $\simu{P}$ can perform operations that violate ``charge'' conservation. Our task is to ensure that the greater freedom enjoyed by a dishonest party in the $U$-world does not enhance her ability to cheat successfully.

For this purpose, our argument relies on the concept of the {\em format} of a message exchanged between the parties. In the $U$-world, the format is simply the Hilbert space containing the message. In the protocol $\simu{P}$, the recipient of a message always checks that the format of the message is valid, and aborts the protocol if the message is invalid. A valid message corresponds to one that could have been sent in the $I$-world, while a message is invalid only if the sender violated the local conservation of ``charge'' before sending it. Thus, a message that upon receipt is found to be in the proper format could have been sent by a party who performed a charge-conserving local operation --- in effect the sender is unable to play a charge nonconserving strategy without being detected.
Since effective charge conservation is enforced by halting the protocol when a charge nonconservation is detected, it will be essential for our argument to consider games that can be aborted at any stage by either party. A cheating strategy for the $I$-world protocol $P$ and the corresponding cheating strategy for its $U$-world counterpart $\simu{P}$ will cause the game to halt prematurely with the same probability, as well as produce the same probability distribution of outcomes in the event that the game ends normally, without being aborted.

\subsection{Superselection rules and charges}

Before proceeding to our proof, we should recall the properties of superselection rules and charges that will be invoked in the argument. These properties have been explored already in Sec.~\ref{sec:commutants}, for the special case of super-selection sectors labeled by irreducible unitary representations of compact groups. Here we wish to emphasize that some of the same ideas can be extended to a more general setting, and we will indicate how a two-party protocol in which conserved charges are exchanged can be simulated using ordinary qubits. 

In general, a superselection rule is a decomposition of Hilbert space into a direct sum of sectors such that each sector is preserved by the allowed operations. The charge $q$ is a label that distinguishes the distinct sectors, and we may say that the operations allowed by the superselection rule conserve the charge. Thus, the Hilbert space is expressed as 
\begin{eqnarray}
{\cal H}= \bigoplus_q~ {\cal H}_q~,
\end{eqnarray}
and the allowed operations belong to the algebra
\begin{eqnarray}
\bigoplus_q~{\cal L}\left({\cal H}_q\right)~,
\end{eqnarray}
where ${\cal L}\left({\cal H}_q\right)$ denotes linear operators acting on ${\cal H}_q$.

Depending on the particular form of the superselection rule, there are specific rules governing how the charge behaves when a system splits into two subsystems, or when two systems fuse to become a single system. These rules can be encoded in vector spaces $V_c^{a,b}$ defined by
\begin{equation}
{\cal H}_c = \bigoplus_{a,b}~{\cal H}_a\otimes {\cal H}_b\otimes V_c^{a,b}~.
\label{eq:fuse_and_split}
\end{equation}
The space $V_c^{a,b}$ is $n$-dimensional if there are $n$ distinguishable ways that a charge $c$ object can arise when objects with charges $a$ and $b$ fuse. Consistency of eq.~(\ref{eq:fuse_and_split}) with associativity of the tensor product requires the $V_c^{a,b}$'s to obey certain identities, but we will not discuss these further as they will not be needed for our proof.

There is a trivial-charge sector, denoted ${\cal H}_1$, that behaves as the identity under fusion:
\begin{equation}
{\cal H}_c\otimes{\cal H}_1= {\cal H}_c~.
\end{equation}
Furthermore, there is a unique charge $\bar q$, the conjugate of $q$, that can fuse with $q$ to yield the identity:
\begin{equation}
{\cal H}_1= \bigoplus_q~{\cal H}_q\otimes {\cal H}_{\bar q}~.
\end{equation}

Now, in the $I$-world, consider a bipartite system shared by Alice and Bob. The Hilbert space decomposes as
\begin{eqnarray}
\label{eq:the_whole_Hilbert}
&&{\cal H}=\bigoplus_q~ {\cal H}_q ~,\nonumber\\
&&{\cal H}_q=\bigoplus_{q_A,q_B}~ {\cal H}_{A,q_A}\otimes {\cal H}_{B,q_B}\otimes V_q^{q_A,q_B}~,
\end{eqnarray}
where $q$ is the total charge, $q_A$ is the charge of Alice's system, and $q_B$ is the charge of Bob's system. The physical operations, allowed by the superselection rule, conserve the total charge, and hence belong to the algebra 
\begin{equation}
{\cal O}=\bigoplus_q {\cal L}\left({\cal H}_q\right)~.
\end{equation}
The operations Alice can perform, which conserve Alice's charge and act trivially on Bob's system, belong to
\begin{equation}
{\cal A}=\bigoplus_{q,q_A,q_B}{\cal L}\left({\cal H}_{A,q_A}\right)\otimes I_{B,q}^{q_A,q_B}~,
\end{equation}
where $I_{B,q}^{q_A,q_B}$ denotes the identity acting on ${\cal H}_{B,q_B}\otimes V_q^{q_A,q_B}$.
Similarly, the algebra of operations that Bob can perform is
\begin{equation}
{\cal B}=\bigoplus_{q,q_A,q_B}I_{A,q}^{q_A,q_B} \otimes{\cal L}\left({\cal H}_{B,q_B}\right)~,
\end{equation}
where $I_{A,q}^{q_A,q_B}$ denotes the identity acting on ${\cal H}_{A,q_A}\otimes V_q^{q_A,q_B}$.
In contrast, the commutant ${\cal B}'$ of ${\cal B}$, which conserves the total charge and Bob's charge but need not conserve Alice's, is
\begin{eqnarray}
{\cal B}'= \bigoplus_{q,q_B}{\cal L}\left(\bigoplus_{q_A} {\cal H}_{A,q_A}\otimes V_q^{q_A,q_B}\right)\otimes I_{B,q_B}~,
\end{eqnarray}
where $I_{B,q_B}$ is the identity on ${\cal H}_{B,q_B}$, and similarly
\begin{eqnarray}
{\cal A}'= \bigoplus_{q,q_A}I_{A,q_A}\otimes{\cal L}\left(\bigoplus_{q_B} {\cal H}_{B,q_B}\otimes V_q^{q_A,q_B}\right)~.
\end{eqnarray}
Thus ${\cal A}'={\cal B}$ and ${\cal B}'={\cal A}$ if and only if the charges $q_A$ and $q_B$ are perfectly correlated (there is a unique $q_B$ corresponding to each $q_A$ and vice versa). This condition holds, in particular, if the total charge is trivial, in which case our formulas simplify to
\begin{eqnarray}
&&{\cal H}={\cal H}_1=\bigoplus_q {\cal H}_{A,q}\otimes {\cal H}_{B,\bar q}~,\nonumber\\
&&{\cal A}={\cal B}'= \bigoplus_q {\cal L}\left({\cal H}_{A,q}\right)\otimes I_{B,\bar q}~,\nonumber\\
&&{\cal B}={\cal A}'= \bigoplus_q I_{A,q}\otimes {\cal L}\left({\cal H}_{B,\bar q}\right)~.
\end{eqnarray}

\subsection{Simulating charge exchange}
\label{subsec:exchange}
A novelty of a two-party protocol in the $I$-world is that when Alice (for example) sends a message to Bob, she may choose to split the charge she possesses into two parts --- the charge she retains and the charge of the message that she sends. If the total charge is trivial, then the full Hilbert space comprising Alice's system $A$, Bob's system $B$, and the message system $M$ can be expressed as
\begin{equation}
\label{eq:Alice_Bob_message}
{\cal H}_1=\bigoplus_{q_A,q_B,q_M}{\cal H}_{A,q_A}\otimes {\cal H}_{B,q_B}\otimes {\cal H}_{M,q_M}\otimes V_1^{q_A,q_B,q_M}~.
\end{equation}
The isomorphisms 
\begin{equation}
\label{V_isomorph}
V_1^{q_A,q_B,q_M}\cong V_{\bar q_B}^{q_A,q_M}\cong V_{\bar q_A}^{q_B,q_M}
\end{equation}
invite us to interpret eq.~(\ref{eq:Alice_Bob_message}) in complementary ways --- namely, the charge $\bar q_B$ of $AM$ is conjugate to the charge $q_B$ of $B$, and the charge $\bar q_A$ of $BM$ is conjugate to the charge $q_A$ of $A$. Thus, eq.~(\ref{eq:Alice_Bob_message}) describes the splitting of Alice's initial charge $\bar q_B$ into the charge $q_A$ that she retains and the charge $q_M$ of the message, as well as the fusion of the charge $q_M$ of the message with Bob's initial charge $q_B$ to yield Bob's final charge $\bar q_A$. Furthermore, if $V_1^{q_A,q_B,q_M}$ is of dimension greater than one, then a vector in $V_1^{q_A,q_B,q_M}$ describes the particular manner in which Alice performs the splitting, which in turn determines the result of Bob's fusion.

While the information encoded in $V_1^{q_A,q_B,q_M}$ is an intrinsic property in the $I$-world, if we are to simulate the process of charge exchange in the $U$-world, then this information must be carried by ordinary qubits. In such a simulation, the Hilbert space of Alice's system, Bob's system, and the message is expanded to
\begin{equation}
\simu{\HH}=\bigoplus_{q_1,q_2,q_A,q_B, q_M}{\cal H}_{A,q_1}\otimes {\cal H}_{B,q_2}\otimes {\cal H}_{M,q_M}\otimes V_1^{q_A,q_B,q_M}~,
\end{equation}
but where now $V_1^{q_A,q_B,q_M}$ is to be regarded as an explicit part of the message.
If the conditions $q_1=q_A$ and $q_2=q_B$ were imposed, then the ``format'' of this message would coincide perfectly with the information content of a message sent in the $I$-world. But while in the $I$-world these conditions arise from the intrinsic physics of the superselection rule, in the $U$-world they must be imposed by hand through proper design of the protocol.

Thus, in the $U$-world protocol $\simu{P}$ that simulates the $I$-world protocol $P$, we will require the recipient of a message to verify its format --- Alice checks that $q_1=q_A$ and Bob checks that $q_2=q_B$. Of course, at a given stage of the protocol $P$, Alice or Bob might hold a coherent superposition of different charges, even though the total charge is always guaranteed to be trivial. Therefore the verification step in $\simu{P}$ must be performed coherently; Alice, for example, checks that $q_1$ and $q_A$ match without learning the value of $q_1$ or $q_A$. If verification fails, then the message recipient has detected cheating by the other party and aborts the protocol. If verification succeeds, then the message has been projected onto the valid format, and as far as the recipient is concerned, it is just as though the message had been sent in the right format to begin with. 

Whenever Alice cheats in the $U$-world protocol $\simu{P}$ by modifying her charge, she risks detection, and if her cheating is undetected, then her operation is equivalent to a charge-conserving one. Therefore, Alice has an equivalent strategy in the $I$-world protocol $P$, in which she either halts the game herself with some probability before sending her message, or if the game does not halt, performs an operation allowed by the superselection rule. This observation suffices to establish that $\simu{P}$ simulates $P$, and thus that the superselection rule cannot thwart cheating.

To summarize, for the purpose of characterizing Alice's ability to cheat, we are only interested in how Alice's activities will affect Bob's measurements. Although in the $U$-world Alice has the power to violate conservation of ``charge,'' she is unable to fool Bob into accepting a message that is not isomorphic to one that could have been created in the $I$-world. Therefore, Alice's elevated power in the $U$-world gives her no advantage.

\subsection{Definitions}
\begin{figure}[h]
\centering
\begin{picture}(132,80)

\put(0,0){\line(1,0){132}}
\put(0,20){\line(1,0){132}}
\put(0,40){\line(1,0){132}}
\put(0,60){\line(1,0){132}}
\put(0,80){\line(1,0){132}}

\put(0,0){\line(0,1){80}}
\put(60,0){\line(0,1){80}}
\put(96,0){\line(0,1){80}}
\put(132,0){\line(0,1){80}}

\put(96,08){\makebox(0,0){$\longleftarrow$}}
\put(96,28){\makebox(0,0){$\longleftarrow$}}
\put(96,48){\makebox(0,0){$\longrightarrow$}}

\put(30,08){\makebox(0,0){Bob strategy}}
\put(30,28){\makebox(0,0){Alice strategy}}
\put(30,48){\makebox(0,0){Protocol}}

\put(78,68){\makebox(0,0){$I$-world}}
\put(114,68){\makebox(0,0){$U$-world}}

\put(76,08){\makebox(0,0){$B'$}}
\put(114,08){\makebox(0,0){$\simu{B}'$}}
\put(76,28){\makebox(0,0){$A'$}}
\put(114,28){\makebox(0,0){$\simu{A}'$}}
\put(76,48){\makebox(0,0){$P$}}
\put(114,48){\makebox(0,0){$\simu{P}$}}

\end{picture}
\caption{The $U$-world protocol $\simu{P}$ {\em simulates} the $I$-world protocol $P$ if the honest protocols realize the same task, and if for any cheating strategy in $\simu{P}$ there is an equivalent cheating strategy in $P$.}
\label{fig:simulate}
\end{figure}

Having explained the main ideas, we will now present a more formal proof of our result.
To begin, we must define the general notions of ``protocol'' and ``simulation''
in accord with our goals. The definitions are quite natural, but there are some
technicalities that are necessary for the proof to work.

We consider quantum games between two parties, Alice and Bob. We assume that
Alice sends the first message and the players alternate. The {\em protocol} of a game
specifies the total number of messages, their format, the
strategies for honest players, and a way to determine the game outcome. By
``format'' in the $U$-world we mean the Hilbert space $\HH_{M}$ of a given
message. In the $I$-world, we specify the space $\HH_{M,q_{M}}$ for each value of the message charge $q_{M}$.

To define an honest strategy in the $I$-world, we specify for each value of Alice's charge $q_A$ her corresponding space
$\HH_{A,q_{A}}$; likewise, we specify Bob's space $\HH_{B,q_{B}}$ for each $q_{B}$. The game starts with a pure
state
\begin{eqnarray}
|\xi_{A}\rangle\otimes|\xi_{B}\rangle \in \HH_{A,1}\otimes\HH_{B,1},
\end{eqnarray}
where $1$ stands for the trivial charge. If one of the players (say, Alice)
cheats, she may use a different set of private spaces $H_{A,q_{A}}'$, but the
initial state still must be of the form $|\xi_{A}'\rangle\otimes|\xi_{B}\rangle$,
where $|\xi_{A}'\rangle\in\HH_{A,1}'$.

Alice's and Bob's actions in the $k$th step are
described by operators $W_{A_{k}}$, $W_{B_{k}}$. 
The final outcome is determined by a pair of measurements that are performed
independently on Alice's and Bob's subsystems at the end of the game. We are
interested in the joint probability distribution of the measurement results.
However, if one of the players cheats, only the honest player's subsystem is
measured.

For the reasons explained in Sec.~\ref{subsec:overview}, we will assume that the game can be aborted by either player. If the game is aborted, we will not need to keep track of who ends the game or when it ends --- we will only be interested in whether the game ends normally and if so what is the outcome. For this purpose the quantum state can be characterized by a vector
$|\psi\rangle$ such that $\langle\psi|\psi\rangle$ is the probability that the
game has not been aborted. Operations performed by each player may then be described
by contracting maps, i.e., operators $W$ such that $W^{\dagger}W\le I$. We assume that the game is never aborted if both players are honest, so that the
probabilities of different outcomes add up to $1$ in the honest game. If one of the players
cheats, the total probability of all outcomes is generally less than $1$.

Now we define what it means for one protocol to {\em simulate} another (see Fig.~\ref{fig:simulate}):

\medskip
\noindent {\bf Definition} 
{\em A protocol $\simu{P}$ \emph{simulates} the protocol $P$ if the
following conditions are fulfilled:
\begin{enumerate}
\item The honest strategies in $P$ and $\simu{P}$ give rise to the same
probability distribution of the outcomes.
\item For any cheating strategy $\simu{A}'$ by Alice compatible with the
protocol $\simu{P}$ there exists an equivalent strategy $A'$ for the protocol
$P$. (``Equivalent'' means that Bob's measurement result has the same
probability distribution in both cases.)
\item For any  cheating strategy $\simu{B}'$ by Bob compatible with the protocol
$\simu{P}$ there is an equivalent strategy $B'$ for the protocol $P$.
\end{enumerate}}

\noindent Note that when we say that the two cheating strategies are equivalent we mean in particular that the probability that the game ends normally is the same for both strategies.

To better understand our concept of simulation it is very helpful to consider this simple example: Suppose that the message space $\HH_{M}$ of $P$ is embedded in a larger space $\simu{\HH}_{M}$ of $\simu{P}$.
Honest players follow the same strategies in $\simu{P}$ as in $P$, so that condition~1 is obviously
satisfied. However, the players in $\simu{P}$ must be prepared to receive messages that
do not obey the format of $P$, i.e., do not fit into the subspace $\HH_{M}$. In $\simu{P}$ such messages are rejected, and the game is aborted. This
rule prevents a dishonest player from gaining any advantage (relative to simply quitting the game)
by sending an invalid message. More
formally, suppose that Alice cheats using some strategy $\simu{A}'$. In the 
corresponding strategy $A'$, Alice projects her message system $\simu{\HH}_{M}$ onto the subspace
$\HH_{M}$, before sending each message. Thus if the strategy $\simu{A}'$ calls for Alice to apply the operator $\simu{W}_{A_{k}}'$ in the $k$th round, then in the strategy $A'$ Alice applies the contracting map $W_{A_{k}}'=\Pi\simu{W}_{A_{k}}'$, where $\Pi$ is the orthogonal projector onto $\HH_{M}$. The strategies $\simu{A}'$ and $A'$ are equivalent: whenever a message
sent according to $\simu{A}'$ causes Bob to abort the game, the strategy $A'$
requires Alice to abort the game herself. Similarly, given any cheating strategy $\tilde B'$ for Bob in the game $\simu{P}$, there is an equivalent cheating strategy $B'$ in $P$. Thus, conditions 2 and 3 are satisfied, and $\simu{P}$ simulates $P$.

Our analysis of superselection rules in Sec.~\ref{subsec:proof} will be based on a closely related method of simulation.

We also remark that Theorem 1 proved in Sec.~\ref{sec:simulation} can be restated: for a multiparty protocol $P$ in the $G$-invariant world, there is a $U$-world protocol $\simu{P}$ that simulates $P$. In that case, we implicitly adopt a redundant description of the physical states appearing in $P$, admitting fictitious color degrees of freedom. Then $\simu{P}$ is exactly the same protocol as $P$, but with the color now reinterpreted as a physical variable. Similarly, Theorem 3 in Sec.~\ref{subsec:nontrivial_charge} can be stated: any $n$-party $I$-world protocol in which the initial state is a product of $n$ invariant states can be simulated by an $I$-world protocol in which the initial state is a product of $n$ pure states, each with trivial charge.

\subsection{Proof}
\label{subsec:proof}

Our goal is to prove:

\medskip
\noindent{\bf Theorem 4} {\em Let $P$ be a two-party game in the $I$-world, such that both parties hold trivial charges at the beginning of the game. Then there is a $U$-world game $\simu{P}$ that simulates $P$.}
\medskip

\noindent In the proof we construct the $U$-world protocol $\simu{P}$ that simulates the $I$-world protocol $P$, and explain how the cheating strategy $A'$ that is equivalent to $\simu{A}'$ is formulated. We achieve this by applying the procedure for simulating charge exchange in the $U$-world that was described in Sec.~\ref{subsec:exchange}.

Consider the $I$-world protocol $P$. If the total charge is trivial, then the full Hilbert space including Alice's system $A$, Bob's system $B$, and the message $M$ is 
\begin{equation} \label{H}
\HH= \bigoplus_{q_{A},q_{B},q_{M}}
\HH_{A,q_{A}}\otimes \HH_{B,q_{B}}\otimes \HH_{M,q_{M}}\otimes
V_{1}^{q_{A},q_{B},q_{M}}~.
\end{equation}
Without loss of generality we assume that the spaces
$\HH_{A,q_{A}}$, $\HH_{B,q_{B}}$, $\HH_{M,q_{M}}$ are the same in each step of
the protocol. We may also assume that the message is present at the beginning
and at the end of the game and that the initial state has the form
$|\xi_{A}\rangle\otimes|\xi_{B}\rangle\otimes|0\rangle$, where
$|0\rangle\in\HH_{M,1}$.

Each time Alice receives one message and sends another, she applies an operator to $AM$ that preserves Bob's charge $q_B$; this is a contracting map belonging to the algebra
\begin{equation}
\bigoplus_{q_{B}} \LL\left( \bigoplus_{q_{A},q_{M}}
\HH_{A,q_{A}}\otimes \HH_{M,q_{M}}\otimes V_{1}^{q_{A},q_{B},q_{M}} \right).
\end{equation}
Alice's honest strategy consists of a sequence of such
operators --- in the $k$th step she applies an operator $W_{A_{k}}$. Similarly, Bob's honest
strategy is defined by operators $W_{B_{k}}$.

Now consider the $U$-world protocol $\simu{P}$ that simulates $P$. The Hilbert space of $\simu{P}$ is
\begin{equation}
\simu{\HH}=\simu{\HH}_A\otimes \simu{\HH}_B\otimes \simu{\HH}_M~,
\end{equation}
where
\begin{equation}
\begin{array}{c}
\displaystyle
\simu{\HH}_{A}=\bigoplus_{q_{1}}\HH_{A,q_{1}}~, \qquad
\simu{\HH}_{B}=\bigoplus_{q_{2}}\HH_{B,q_{2}}~,
\bigskip\\
\displaystyle
\simu{\HH}_{M}=\bigoplus_{q_{A},q_{B},q_{M}}
\HH_{M,q_{M}}\otimes V_{1}^{q_{A},q_{B},q_{M}}~.
\end{array}
\end{equation}
Thus the space $\HH$ of the protocol $P$ can be
embedded in $\simu{\HH}$ by requiring 
$q_{1}=q_{A}$ and $q_{2}=q_{B}$. In $\simu{P}$, these constraints are
enforced by checks performed by both parties. A dishonest player's
attempt to break the constraints will be detected immediately by the other
party, in which case the game will halt.

Let us describe Alice's honest strategy in $\simu{P}$. When Alice
receives a message, she gains control of the space
$\simu{\HH}_{A}\otimes\simu{\HH}_{M}$. First she verifies that $q_{1}=q_{A}$ (without determining the value of $q_1$ or $q_A$); if verification fails, she aborts the game. Thus Alice
effectively projects her input state onto the subspace
\begin{equation}
\HH_{AM}=\bigoplus_{q_A,q_B,q_M}{\cal H}_{A,q_A}\otimes {\cal H}_{M,q_M}\otimes V_1^{q_A, q_B, q_M}\subseteq\simu{\HH}_{A}\otimes\simu{\HH}_{M}~.
\end{equation}
Then she applies the operator $W_{A_{k}}$ (from the
protocol $P$), which acts on $\HH_{AM}$ and preserves $q_{B}$. Thus Alice's strategy is defined by the contracting maps
\begin{eqnarray}
\simu{W}_{A_{k}}=FW_{A_{k}}F^{\dagger},
\end{eqnarray}
where $F$ denotes the embedding
$\HH_{AM}\to\simu{\HH}_{A}\otimes\simu{\HH}_{M}$. Bob's honest strategy is
defined similarly. 

If both players play the game $\simu{P}$ honestly, then the verification always succeeds and the conditions $q_{1}=q_{A}$ and $q_{2}=q_{B}$ are maintained throughout the game. Thus the honest strategies for $\simu{P}$ and $P$ are clearly equivalent. Note that in $\simu{P}$ some information is encoded redundantly --- for example Alice can access the value of $q_A$ by examining either the charge label of ${\cal H}_{A,q_A}$ or one of the slots of the tensor $V_1^{q_A,q_B,q_M}$; similarly $q_M$ is encoded both in ${\cal H}_{M,q_M}$ and in $V_1^{q_A,q_B,q_M}$. However, this redundancy has no deleterious effect on the fidelity of the simulation.

Now suppose that Alice cheats in the game $\simu{P}$. Then she may use an arbitrary Hilbert
space $\simu{\HH}_{A}'$ and operators $\simu{W}_{A_{k}}'$ acting on
\begin{eqnarray}
\label{tildeP_cheating}
&&\simu{\HH}_{AM}' = \simu{\HH}_{A}'\otimes\simu{\HH}_{M}\nonumber\\
&&=\simu{\HH}_{A}'\otimes\Big(\bigoplus_{q_{A},q_{B},q_{M}}
\HH_{M,q_{M}}\otimes V_{1}^{q_{A},q_{B},q_{M}} ~\Big).
\end{eqnarray}
In particular, when Alice cheats her action on the message need not respect the condition $q_B=q_2$. To prove the theorem, we are to define an equivalent cheating strategy for the game $P$. 

When Alice cheats in $P$, she uses an arbitrary Hilbert space $\HH_{A,q_A}'$ for each value of her charge $q_A$, 
and she applies  operators $W_{A_k}'$ that conserve Bob's charge $q_B$ to the space
\begin{equation}
\HH_{AM}' = \bigoplus_{q_{A},q_{B},q_{M}}
\HH_{A,q_{A}}'\otimes \HH_{M,q_{M}}\otimes
V_{1}^{q_{A},q_{B},q_{M}} ~.
\end{equation}
The spaces $\simu{\HH}_{AM}'$ and $\HH_{AM}'$ seem to be distinct --- in $\HH_{AM}'$ the charge label carried by ${\cal H}_{A,q_A}'$ matches the label in one of the slots of $V_{1}^{q_{A},q_{B},q_{M}}$, while in $\simu{\HH}_{AM}'$ there is no such correlation. However, in the $U$-world the variable $q_A$ would be encoded redundantly if it appeared in both ${\cal H}_{A,q_A}'$ and $V_{1}^{q_{A},q_{B},q_{M}}$, and it is not necessary to adopt this redundant encoding in order to emulate the physics of the $I$-world. Instead, let us specify $\HH_{A,q_A}'=\simu{\HH}_{A}'$ for each $q_A$ --- then $\HH_{AM}'$ and $\simu{\HH}_{AM}'$ are of the same form, but where it is understood in eq.~(\ref{tildeP_cheating}) that the information about the charge $q_A$ is carried only by  $V_1^{q_A,q_B,q_M}$. With this choice Alice's operator $\simu{W}_{A_{k}}'$ in $\simu{P}$ and her operator $W_{A_k}'$ in $P$ act on isomorphic spaces;  however $W_{A_k}'$ must conserve Bob's charge $q_B$, while $\simu{W}_{A_{k}}'$ need not conserve charge. 

Therefore, we define the corresponding cheating strategy in $P$ by specifying
\begin{equation}
W_{A_k}'=\sum_{q_{B}} \Pi_{q_{B}}\simu{W}_{A_{k}}'\Pi_{q_{B}}~,
\end{equation}
where $\Pi_{q_{B}}$ is the projector onto the subspace with the given value of
$q_{B}$. That is, $\Pi_{q_{B}}$ projects $\simu{{\cal H}}_M$ onto the space in which $V_1^{q_A,q_B,q_M}$ has the value $q_B$ in the appropriate slot. The contracting map  $W_{A_k}'$ preserves $q_{B}$ and therefore is admissible in the protocol $P$. Applying this $W_{A_k}'$ causes {\em Alice} to abort the game $P$ in
the case where $q_{B}$ would change in the game $\simu{P}$. But in that case the new value of $q_{B}$ would not match Bob's variable $q_{2}$; therefore {\em Bob} would reject Alice's message and abort the game $\simu{P}$. Hence the two games $P$ and $\simu{P}$ are aborted with the same probability; furthermore, the final state that Bob measures in $\simu{P}$, if $\simu{P}$ does not abort, is identical to the final state that Bob measures in $P$, if $P$ does not abort. Therefore, when Alice cheats, Bob's measurement outcome has the same probability distribution in $\simu{P}$ as in $P$. The same is true for Alice's measurement when Bob cheats. Therefore, $\simu{P}$ simulates $P$, which completes the proof of Theorem 4.

\section{Conclusions}
\label{sec:conclusions}

Recent progress in the theory of quantum computation and quantum cryptography highlights the importance of adopting a computational model compatible with fundamental physics --- tasks that would be impossible in a classical world may be physically realizable because Nature is quantum mechanical. Further refinements of the model could lead to further insights regarding what information-processing tasks are achievable. Therefore, as Popescu \cite{popescu} emphasized, the impact of superselection rules on the security of quantum protocols is of considerable potential interest. However, our disappointing conclusion is that superselection rules cannot foil a cheater who has unlimited quantum-computational power.

Contemplating this issue has led us to consider how physics in the invariant world can simulate physics in the unrestricted world, and vice versa. We feel that the simulation schemes we have devised offer fruitful insights into the physical meaning of superselection rules.

Our results do not address whether the security of protocols with more than two parties can be enhanced by superselection rules that do not arise from compact symmetry groups. New issues arise in this setting, because of the nontrivial braiding properties of nonabelian anyons. For example, in the case of three parties (Alice, Bob, and Charlie), Alice can split her charge into two parts, and send one part on a voyage that circles Bob's lab and then returns to Alice's lab. This action can induce a change in the charge held by Alice, accompanied by a compensating change in the total charge held by Bob and Charlie, even though the local charge in Bob's lab, and in Charlie's, is unaltered. Though strictly speaking Alice's operation is not ``local,'' she can carry it out surreptitiously, without any cooperation from Bob and Charlie. Such new possibilities enhance the potential power of cheaters, but may also provide the honest parties with new methods for detecting cheating. Addressing the security of multiparty quantum protocols subject to general superselection rules will require different methods than we have used in this paper, and might provide further enlightenment concerning the physics of nonabelian anyons.

\acknowledgments
We thank Stephen Bartlett, Michael Ben-Or, and Sandu Popescu for discussions. This work has been supported in part by the Department of Energy under Grant No. DE-FG03-92-ER40701, by the National Science Foundation under Grant No. EIA-0086038, and by the Caltech MURI Center for Quantum Networks under ARO Grant No. DAAD19-00-1-0374.

\end{document}